%% file: bare_jrnl_new_sample4.tex
\documentclass[lettersize,journal]{IEEEtran}

\usepackage{cite}
\usepackage{amsthm}

\usepackage{amsmath}
\usepackage{amsfonts}
\usepackage{lipsum} % Example package -- can be removed
\usepackage[utf8]{inputenc}
\usepackage[hidelinks]{hyperref}
 \usepackage{ulem}
 \usepackage{subfigure}
 \usepackage{caption}
 \usepackage{comment}
 \usepackage{url}
\usepackage{multirow}
\usepackage{graphicx}
 \usepackage{booktabs} %用于加粗表格线条
\usepackage[ruled]{algorithm}
 \usepackage{algorithmicx}
 \usepackage[noend]{algpseudocode}
 \usepackage[para]{threeparttable} % 表格注释尽量一行显示
\usepackage{listings}
\usepackage{tabularx,tabulary}

\usepackage{lipsum} % Example package -- can be removed
\usepackage[utf8]{inputenc}
 \usepackage{hyperref}

\begin{document}
%%%% 3. TITLE %%%%
\title{Optimized Vectorization Implementation of CRYSTALS-Dilithium}
%%%% 4. AUTHOR, INSTITUTE %%%%
\author{Jieyu Zheng, Haoliang Zhu, Zhenyu Song, Zheng Wang, Yunlei Zhao}
        % <-this % stops a space

\markboth{Journal of \LaTeX\ Class Files,~Vol.~14, No.~8, August~2021}%
{Shell \MakeLowercase{\textit{et al.}}: A Sample Article Using IEEEtran.cls for IEEE Journals}


\maketitle
%%%% 5. ABSTRACT %%%%
\begin{abstract}
CRYSTALS-Dilithium  is a lattice-based signature scheme to be  standardized by NIST as the primary post-quantum signature algorithm. In this work, we make a thorough study of optimizing the implementations of Dilithium by utilizing the Advanced Vector Extension (AVX) instructions, specifically AVX2 and the latest AVX-512. We first  present an improved parallel small polynomial multiplication with tailored early evaluation (PSPM-TEE)  to further speed up the signing procedure. Our PSPM algorithm outperform the NTT by 47\%-66\% in AVX2 and AVX-512 implementation. We then present a tailored reduction method that is simpler and  faster than Montgomery reduction. We minimize the CPU cycles of tailored reduction AVX-512 implementation by using AVX-512IFMA. Finally, we propose a fully and highly vectorized implementation of Dilithium using AVX-512. This is achieved by carefully   vectorizing  most of Dilithium functions with the AVX-512 instructions  in order to improve efficiency  both for time and for space simultaneously. With all the optimization efforts,  our AVX-512 implementation improves the performance by 43.2\%/39.3\%/45.6\% in key generation, 36.6\%/41.6\%/43.7\% in signing,  and 45.3\%/46.5\%/47.4\% in verification for the parameter sets of  Dilithium2/3/5  respectively.  To the best of our knowledge, our AVX-512 implementation has the best performance for Dilithium on the Intel x86-64 CPU platform to date.
\end{abstract}
%%%% 6. KEYWORDS %%%%

%%%% 7. PAPER CONTENT %%%%
%%% Introduction %%%%%
\section{Introduction}
\IEEEPARstart{W}{ith} the popularity of authentication and non-repudiation, it is more common to construct digital signatures using asymmetric cryptographic techniques. Currently, millions of web servers use digital signatures as part of Transport Level Security (TLS) \cite{itu2000information,dierks2008transport,rescorla2018transport}, which allows users to verify the server’s identity. Both hardware and software vendors rely on digital signatures to guarantee entity integrity. Digital signatures are also essential for cybersecurity infrastructure. Most of the current digital signatures are implemented based on Rivest-Shamir-Adleman (RSA) \cite{rivest1978method}, Elliptic Curve Cryptography (ECC), or Digital Signature Algorithm (DSA).

However, in the era of continuous development of quantum computers, traditional public key cryptography and DSA appear to be in jeopardy. Using Shor’s algorithm \cite{shor1999polynomial}, an attacker with a powerful quantum computer can obtain the corresponding private key in polynomial time by analyzing the public key of RSA or ECC. The National Institute of Standards and Technology (NIST) proposed in \cite{mckay2016report} that by 2030 an RSA 2048-bit key may be broken by a quantum computer within a few hours. As a result, NIST has launched a competition to solicit standard algorithms for PQC, including soliciting and evaluating quantum-resistant secure digital signature algorithms.
On July 5th 2022, NIST announced the first algorithms to be  standardized. There are three signature schemes  selected: CRYSTALS-Dilithium, FALCON and SPHINICS+ \cite{alagic2022status}, among them CRYSTALS-Dilithium is recommended by NIST as the primary signature algorithm to be used. NIST recently released three draft standards. FIPS 204 \cite{dilidraft2023} is among these drafts and pertains to CRYSTALS-Dilithium.  CRYSTALS-Dilithium is a digital signature scheme based on lattice theory, whose security is based on the Module Learning With Errors (MLWE) \cite{DBLP:journals/dcc/LangloisS15} and the Module Short Integer Solution (MSIS) \cite{ajtai1996generating} problems. The majority of Dilithium's operations rely on cyclotomic polynomial ring arithmetic, and it leverages the Number Theoretic Transform (NTT) as a common technique for accelerating polynomial multiplication. The Dilithium scheme adopts the Fiat-Shamir with Aborts structure \cite{lyubashevsky2009fiat}, resulting in a signature process that carefully scrutinizes and rejects sampling through a series of conditional checks. This rigorous process ensures that the generated signature does not divulge any private key information.

NIST chose the 64-bit Intel architecture (i.e. x86-64) as the main benchmarking platform of NIST PQC candidates. Advanced Vector Extension (AVX) is Intel x86-64 instruction set architecture \cite{ lomont2011introduction}. The first AVX instruction was proposed by Intel in 2008. AVX-512 is the newest version of Intel Advanced Vector eXtensions \cite{avx51217}. It has 32 512-bit vector registers called \texttt{zmm} registers. The vector registers are partitioned into distinct data lanes, allowing instructions to be executed concurrently within each lane. This parallel processing technique is referred to as Single Instruction Multiple Data (SIMD). The AVX-512 instruction set excels at accelerating non-sequential processes and delivers optimal performance compared to all other Intel SIMD instruction sets. AVX-512 offers a range of permutation instructions and masked load/stores, which are particularly efficient for implementing hash functions, NTT, and rejection sampling. Additionally, AVX-512IFMA has the potential to significantly accelerate multiply and add operations.
\paragraph{Related Work} Dilithium's optimization efforts encompass both software and hardware aspects. However, this paper places its emphasis on the software-optimized implementation of Dilithium. The basic software implementation is the C \texttt{REF} implementation that the CRYSTALS team submitted to NIST \cite{avanzi2022dilithium}. However, the C \texttt{REF} implementation is not optimized and has lower efficiency. Additionally, the CRYSTALS team provided a faster AVX2 optimized version \cite{avanzi2022dilithium} on x86-64 CPUs. Recently, software optimization studies mainly focus on CPU/GPU environments and embedded systems like ARM. Ravi et al.
%\cite{DBLP:conf/cardis/RaviGCB19} proposed an {early-evaluation} method of conditional checks to speed up the signing procedure. The target platforms of this work are Intel Core i5-4460 and ARM Cortex-M4. 
\cite{ DBLP:journals/tches/GreconiciKS21} presented a signed polynomial representation implementation for Cortex-M4 and proposed various stack consumptions and speed trade-offs for the signing procedure. Kim et al.  \cite{kim2022crystals} presented a method for designing the NTT multiplications of CRYSTALS-Dilithium using advanced SIMD instructions and vector registers.    “Asymmetric multiplication” for matrix-to-vector polynomial multiplication
 was introduced in \cite{DBLP:journals/tches/BeckerHKYY22}. Abdulrahman et al. \cite{DBLP:conf/acns/AbdulrahmanHKS22} proposed to switch to a smaller prime modulus for small polynomial multiplication in the signing procedure of Dilithium. \cite{ bradburyfast} presented optimizations of Dilithium on IBM z15 architecture, and  mentioned that employing some optimization methods with advanced instruction sets like AVX-512 as future work. Zheng et al. \cite{DBLP:conf/acsac/ZhengHSXZ22} presented a parallel small polynomial multiplication (PSPM) algorithm that can fastly compute small vector polynomial multiplication in Dilithium, based on which the C and ARM Neon implementations were proposed.

For AVX-512 implementations of PQC algorithms, some arithmetics  like large integer multiplication, Montgomery multiplication, and NTT AVX-512 implementation have received researchers' attention \cite{bos2014montgomery,gueron2016speeding,orisaka2018finite,edamatsu2018acceleration,takahashi2022implementation,robert2022faster}. Cheng et al. \cite{DBLP:journals/tches/ChengFGR22} proposed a highly vectorized implementation for SIKE. \cite{DBLP:journals/tches/ChengFGRR21} presented an implementation using AVX-512 to batch CSIDH group actions.   \cite{alter2021optimizing} presented an implementation using AVX-512 for SPHINCS+.
  Cabral et al. \cite{cabral2018implementation} presented an optimized AVX-512 implementation for SHA-3 family.
Duowei Lei et al. \cite{lei2023faster} present parallel polynomial sampling and arithmetic implementation to speed up Dilithium scheme. 
% Up till now, to the best of our knowledge, there is still  no AVX-512 optimized implementation for Dilithium. It's also interesting to investigate whether the current state-of-art of  AVX2 implementations for Dilithium can be further improved. 
\paragraph{Contributions} This paper enhances the parallel small polynomial multiplication, as previously seen in ACSAC 2022 \cite{DBLP:conf/acsac/ZhengHSXZ22}, by introducing a tailored early evaluation approach. We proceed to implement Dilithium across all security levels, utilizing SIMD instruction sets on x86-64 CPUs, consequently establishing a new speed record on this platform. Our contributions can be summarized as follows.

\begin{enumerate}

\item We introduce an enhanced parallel small polynomial multiplication with tailored early evaluation (PSPM-TEE) to expedite the signing process. PSPM-TEE is implemented using C, AVX2, and AVX-512 instructions. Notably, PSPM-TEE surpasses NTT in performance across all three implementations.

 \item We introduce a tailored reduction method that outperforms Montgomery reduction. We apply it to the first level of \texttt{NTT}$(t_0)$, \texttt{NTT}$(t_1)$ for Dilithium2/3/5 and \texttt{NTT}$(y)$ for Dilithium2.

 \item We introduce an optimized implementation of the tailored reduction, requiring only two instructions and leveraging AVX-512IFMA. This yields a reduction of one instruction and two cycle counts compared to the AVX-512F implementation. When compared to Montgomery reduction, the tailored reduction with AVX-512IFMA demonstrates superior efficiency, saving up to two instructions and six cycle counts.

% \item We present an optimized AVX2 implementation of
% Dilithium by integrating our improved PSPM-TEE, tailored reduction, and lazy reduction techniques. Our
% optimized AVX2 implementation exhibits a speedup of
% 3\%-8\% compared with the state-of-the-art of Dilithium
% AVX2 software.

  \item %We profile the Dilithium software and analyze performance bottlenecks. 
   % We propose a fully and highly vectorized implementation of Dilithium using AVX-512.  We carefully  vectorize most of Dilithium functions, especially performance bottlenecks including $\texttt{NTT}$, $\texttt{NTT}^{-1}$, Montgomery reduction, hashing, and parallel reject sampling.  In particular,  we present a space-efficient implementation of parallel rejection sampling using AVX-512 without a big precomputation table, as otherwise, the space consumption is infeasible by applying the implementation method using AVX2. With all these optimization efforts,  our AVX-512 implementation improves the performance by 43.2\%/39.3\%/45.6\%in key generation,  36.6\%/41.6\%/43.7\% in signing,  and 45.3\%/46.5\%/47.4\% in verification for the parameter sets of  Dilithium2/3/5  respectively.  To the best of our knowledge, our AVX-512 implementation achieves the best performance for Dilithium on the Intel x86-64 CPU platform thus far.

   We propose a fully and highly vectorized implementation of Dilithium utilizing AVX-512. We meticulously vectorize a majority of Dilithium functions, focusing on performance bottlenecks such as $\texttt{NTT}$, $\texttt{NTT}^{-1}$, Montgomery reduction, hashing, and parallel reject sampling. Notably, we present an efficient implementation of parallel rejection sampling using AVX-512, eliminating the need for a large precomputation table. Through these optimization efforts, our AVX-512 implementation achieves remarkable performance improvements of 43.2\%/39.3\%/45.6\% in key generation, 36.6\%/41.6\%/43.7\% in signing, and 45.3\%/46.5\%/47.4\% in verification across the parameter sets of Dilithium2/3/5, respectively. To the best of our knowledge, our AVX-512 implementation achieves the best performance for Dilithium on the Intel x86-64 CPU platform thus far.

\end{enumerate}
\paragraph{Code} We will later open source our code.
\paragraph{Structure of this paper} This paper is organized as follows. Section \ref{sec:pre} reviews some preliminaries. Section \ref{sec:improved PSPM} presents an improved PSPM with early evaluation. Section \ref{sec:Tailored reduction} introduces the proposed special Tailored reduction. Section \ref{sec:implem} deals with the AVX-512 implementation of Dilithium and presents various  optimization strategies. In Section \ref{sec:experi} we go through the performance results and comparison.

%%% Preliminaries %%%%%
\section{Preliminaries}
\label{sec:pre}

\subsection{Notation}
We denote polynomials by lowercase Latin letter $c$ (the coefficient of a polynomial is  $c_i$, which represents the $i$-th element of $c$), vectors of polynomials by bold lowercase letter $\mathbf{t}$, and matrices by bold upper case letter $\mathbf{A}$. If they are transformed to NTT-domain, then we add a hat to make a tag, e.g., $\hat{c}, \hat{\mathbf{t}}$ and $\hat{\mathbf{A}}$.

Let $\mathbb{Z}_{q} \overset{\text{def}}{=} \mathbb{Z}/q\mathbb{Z}$, $\mathcal{R} \overset{\text{def}}{=} \mathbb{Z}[x]/(x^{n}+1)$, and $\mathcal{R}_{q} \overset{\text{def}}{=} \mathbb{Z}_{q}[x]/(x^{n}+1)$. Element $a_i \in \mathbb{Z}_{q}$ will be represented by one element in $\{-\frac{q-1}{2}, \cdots, 0, \cdots, \frac{q-1}{2}\}$. Polynomial $a \in \mathcal{R}_{q}$ can be represented by $a = \sum_{i=0}^{n-1}a_i \cdot x^n$, where $a_i \in \mathbb{Z}_{q}$.

The operator $\circ$ denotes coefficient-wise multiplication. The operator $||$ concatenates two inputs into a byte stream. For $a_i \in \mathbb{Z}_{q}$, $||a_i||_{\infty}$ denotes $|a_i \mod^{\pm} q|$ (the absolute value of $(a_i \mod^{\pm} q)$). For a finite set $S$ or a distribution $D$, $x \leftarrow S$ denotes random sampling of an element from the set $S$, and $x \leftarrow D$ denotes sampling $x$ according to distribution $D$.  $\lfloor z \rfloor$ means rounding down $z$ and $\lceil z \rfloor$ means rounding to the nearest integer of $z$.\\

\subsection{CRYSTALS-Dilithium Signature Scheme}\label{dili}

CRYSTALS-Dilithium is a post-quantum digital signature algorithm based on the hardness of MSIS and MLWE lattice problems. ML-DSA is derived from CRYSTALS-Dilithium. Algorithm \ref{algo-keygen}, \ref{algo-sign}, and \ref{algo-verify} specify the ML-DSA key generation, signature generation, and signature verification, respectively. The polynomial ring in ML-DSA is $\mathbb{Z}_q[x] /\left(x^n+1\right)$, where $n = 256, q = 8380417$.

The function $\mathsf{NumberOfOne}$ means to count the number of 1's in a vector of polynomials. For the details about the seed expansion functions $\mathsf{ExpandA}$, $\mathsf{ExpandS}$ and $\mathsf{ExpandMask}$, the rounding functions $\mathsf{Power2Round}$, $\mathsf{HighBits}$, $\mathsf{LowBits}$ and $\mathsf{Decompose}$, and the hint functions $\mathsf{MakeHint}$ and $\mathsf{UseHint}$, the generating $c$ polynomial function $\mathsf{SampleInBall}$, the reader can  refer to the Dilithium standard draft \cite{dilidraft2023}.
\input{keygen}

\subsection{Hashing}
The hash functions are two eXtendable Output Functions (XOF), namely \texttt{SHAKE-256} and \texttt{SHAKE-128} \cite{dworkin2015sha}. XOF maps an arbitrary-length bit string to a string of infinitely many bits. These XOF functions are mainly used for generating random bytes of \texttt{SHAKE-128} to sample matrix $\textbf{A}$ and for generating random bytes of \texttt{SHAKE-256} to sample $\textbf{s}$, $\textbf{e}$ and $\textbf{y}$.
\input{sign}
\input{verify}
\subsection{Number Theoretical Transform}
\label{subsec-NTT}
Polynomial multiplications are one of the most expensive parts in massive lattice-based cryptographic schemes. The commonly used technique to accelerate computation is the number theoretic transform (NTT). In Dilithium, the modulus $q$ is chosen so that $q\equiv 1~(\bmod~2n)$ and thus there exists a primitive $2n$-th root of unity in $\mathbb{Z}_q$. Concretely, the recommended parameter setting is $q=8380417,~n=256$ for the sake of security, and the expected primitive 512-th root of unity is $r=1753$. The NTT algorithm maps $\mathbf{f}=f_0+f_1x+\cdots+f_{255}x^{255}\in\mathbb{Z}_q[x]/(x^{256}+1)$ to
\begin{equation*}
    \begin{split}
        &(\mathbf{f} \bmod \mathbb{Z}_q/(x^{128}-r^{128}),~ \mathbf{f} \bmod \mathbb{Z}_q/(x^{128}+r^{128}))\\
        =&((f_0+r^{128}f_{128})+\cdots+(f_{127}+r^{128}f_{255})x^{127},\\
        &(f_0-r^{128}f_{128})+\cdots+(f_{127}-r^{128}f_{255})x^{127})\\
        \in &\mathbb{Z}_q[x]/(x^{128}-r^{128})\times\mathbb{Z}_q[x]/(x^{128}+r^{128})
    \end{split}
\end{equation*}
using FFT trick \cite{DBLP:journals/iacr/Seiler18}. We call this transformation forward NTT (denoted as \texttt{NTT} from here on). To transform back from the NTT domain to the regular domain, the inverse NTT (denoted as $\texttt{NTT}^{-1}$) is computed. By recursively applying this, $\mathbf{f}$ is transformed into its NTT form

\begin{center}
    $\texttt{NTT}(\mathbf{f})=\hat{\mathbf{f}}=(\hat{f}_0,\cdots,\hat{f}_{255})\in \mathbb{Z}_q^{256}$
    $\text{where}\ \hat{f}_i= \mathbf{f} \bmod (x-r^{2i-1})=f(r^{2i-1}),~i=1,\cdots,255$
\end{center}
Since the NTT transform is an isomorphism, we have $$\mathbf{f}\circ\mathbf{g}=\texttt{NTT}^{-1}(\texttt{NTT}(\mathbf{f})\circ\texttt{NTT}(\mathbf{g}))$$

\captionsetup[figure]{labelfont={bf}}
\begin{figure}[htbp]
    \centering
    \subfigure[Cooley-Tukey butterfly]{
        \includegraphics[width=2.6in]{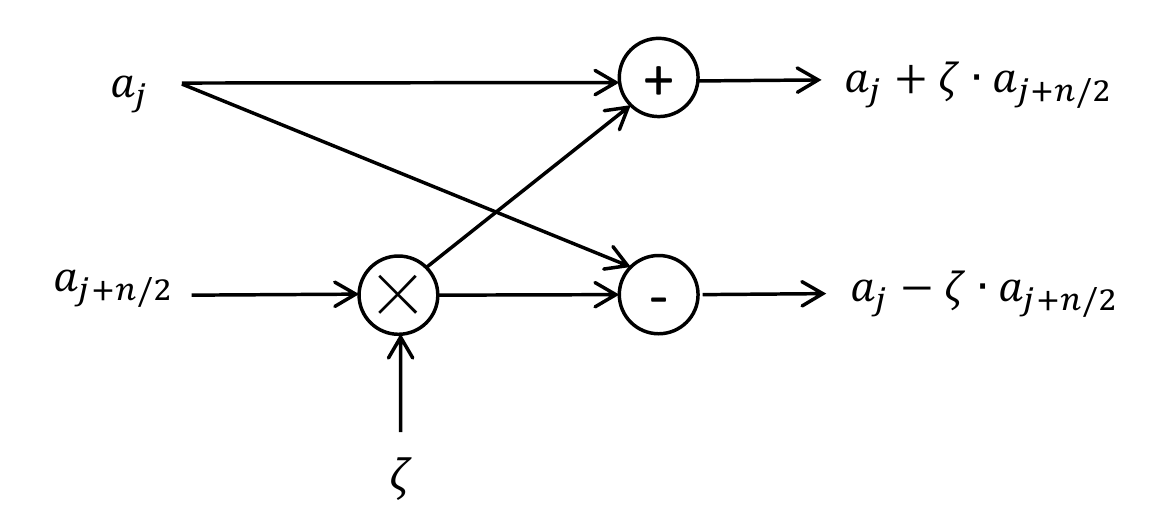}
    }
    \subfigure[Gentleman-Sande butterfly]{
	\includegraphics[width=2.4in]{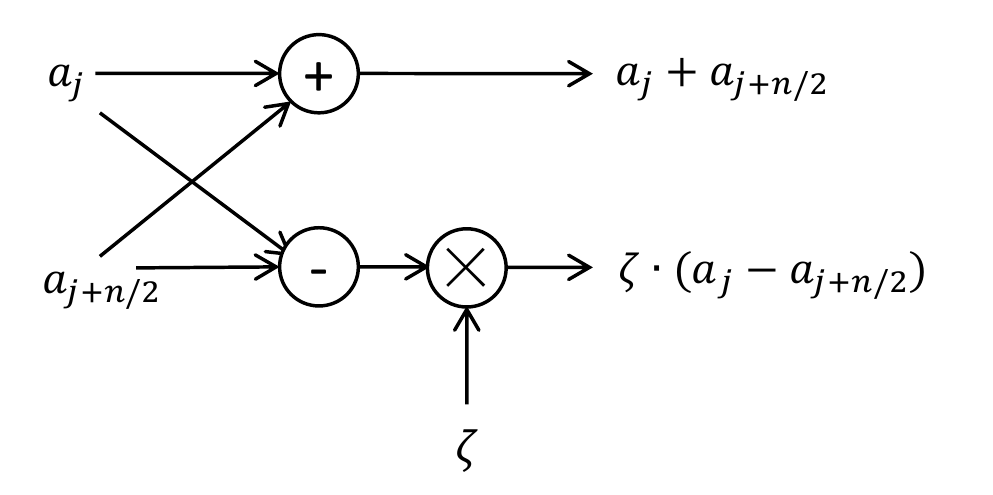}
    }
    \caption{Butterfly diagrams}
    \label{fig:butterfly}
\end{figure}

Note that the direct output of $\texttt{NTT}/\texttt{NTT}^{-1}$ may not result in the natural order as presented, but in a ``bit-reversed'' order. However, each polynomial undergoes two times of bit reversal during NTT  multiplication, one in \texttt{NTT} and one in $\texttt{NTT}^{-1}$, so the result finally turns out in the expected natural order. The core operation to split polynomial $\mathbb{Z}_q[x]/(x^{256}+1)$ to polynomial $\mathbb{Z}_q[x]/(x^{128}-r^{128})$ and $\mathbb{Z}_q[x]/(x^{128}-r^{128})$ is Cooley-Tukey (CT) butterfly \cite{cooley1965algorithm}. The \texttt{NTT} performs $128$ CT butterflies to pairs of coefficients in every iteration of splitting. Each iteration is referred to as a level. Figure \ref{fig:butterfly}(a) depicts the CT butterfly. One might invert the FFT trick using Gentleman-Sande (GS) butterfly \cite{gentleman1966fast}. Figure \ref{fig:butterfly}(b) depicts the GS butterfly.
\input{parallel}

\subsection{Parallel Small Polynomial Multiplication}
As we shall see previously in Section \ref{dili}, one distinctive feature of the polynomial multiplication operations in Dilithium is that many of the time, one of the two multiplicands involved, namely $c \in B_\tau$, has exactly $\tau$ coefficients from 1, -1, the rest being 0. Multiplication by 1 or -1 can be reduced to an addition or subtraction with a sign-based conditional judgment. This is an optimized work presented in  \cite{DBLP:conf/acsac/ZhengHSXZ22}. Algorithm \ref{algo-parallel} is the parallel small polynomial multiplication (PSPM) algorithm, and one single call can compute several products of $c$ and small polynomials, it can speed up the signing and verification of Dilithium. We call lines 1-7 of pseudocode in Algorithm \ref{algo-parallel} \textit{preparing} process, lines 9-21 \textit{evaluating} process.

\subsection{AVX-512 Instruction Set}

Intel Advanced Vector Extensions 512 (AVX-512) is the set of Intel's latest x86-64 vector instructions. AVX-512 adopts the SIMD vectorization parallel approach. Unlike the previous AVX2 instruction set, the size of the vector register is first expanded to 512 bits, and the number of vector registers is also increased from the previous 16 to 32 vector registers (\texttt{zmm0-zmm31}). The AVX-512 vector registers can store more values, and reduce the number of loads from memory to vector registers. In particular, there are eight mask registers in AVX-512 (\texttt{k0-k7}). The mask registers can be used to store the comparison results of two vector registers, enabling more comparison instructions in AVX-512. The mask register can be used for ``maskmov'' type instructions for masking load and store. Generally, we use this type of instructions to select the vector data lane within \texttt{zmm} registers  we need to load or store. AVX-512 has many permutation instructions for adjusting the position of 16-bit, 32-bit, and 64-bit words residing in a \texttt{zmm} register. Such instructions are very important for implementing rejection sampling,  $\texttt{NTT}$ and $\texttt{NTT}^{-1}$, as we shall see. AVX-512F is a vector extension of the x86 instruction set architecture (ISA) that provides 512-bit vector operations, allowing the execution of up to 16 double-precision floating-point or 32 single-precision floating-point operations per cycle. AVX-512F also includes new instructions for integer operations, gather and scatter instructions, and support for masked operations, which allows operations to be selectively applied to vector elements. AVX-512IFMA is an extension to AVX-512F that provides instructions for integer multiplication using the Fused Multiply-Add (FMA) technique, which can perform two multiply-add operations in a single instruction. AVX-512IFMA provides two new IFMA instructions for 52-bit integer \texttt{vpmadd52luq} and \texttt{vpmadd52huq}.

\section{PSPM with Tailored  Early Evaluation (PSPM-TEE) }
\label{sec:improved PSPM}
The signing procedure employs conditional checks for the infinity norm $\mathbf{z}$, $\mathbf{r}_0$, and $c\mathbf{t}_{0}$ to perform rejection sampling.
Since these checks are performed over single coefficients, it is not necessary to compute all the polynomials of the vector. Instead, one polynomial is computed and checked immediately. If the check fails, further computation is unnecessary, saving significant computation time. The probability that $\|\mathbf{z}\|_{\infty}<\gamma_1-\beta$ is
$
\left(\frac{2\left(\gamma_1-\beta\right)-1}{2 \gamma_1-1}\right)^{256 \cdot \ell}=\left(1-\frac{\beta}{\gamma_1-1 / 2}\right)^{\ell n} \approx \mathrm{e}^{-256 \cdot \beta \ell / \gamma_1},
$
and the probability of $\mathbf{r}_0$ in the good range is $
\left(\frac{2\left(\gamma_2-\beta\right)-1}{2 \gamma_2}\right)^{256 \cdot k} \approx \mathrm{e}^{-256 \cdot \beta k / \gamma_2}.
$ 
 It is worth noting that the majority of loop repetitions occur due to the infinity checks of $\mathbf{z}$ and $\mathbf{r}_0$. Therefore, we will only consider the probabilities of these two vectors. In previous implementations, the infinite norm of the vector $\mathbf{z}$ was first evaluated, followed by the evaluation of the infinite norm of vector $\mathbf{r}_0$. In this paper, for the first time, we propose to adjust the order of evaluation of vector $\mathbf{z}$ and vector $\mathbf{r}_0$ based on the different rejection probabilities of vector $\mathbf{z}$ and vector $\mathbf{r}_0$ for different parameters of Dilithium. We can compute the probabilities of the two conditional checks for three parameter sets. As shown in Table \ref{Table:rep}, the probability of vector $\mathbf{z}$ falling within a good range is always greater than the probability of vector $\mathbf{r}_0$. Hence, checking $\mathbf{r}_0$ prior to checking $\mathbf{z}$ can result in a faster signature procedure since repetition is more likely to occur after checking $\mathbf{r}_0$ and the computation of $\mathbf{z}$ can be saved. We tested the performance of the Dilithium C \texttt{REF} implementation between checking $\mathbf{r}_0$ first and checking $\mathbf{z}$ first. We observe that checking $\mathbf{r}_0$ before checking $\mathbf{z}$ results in a 2\% to 3\% improvement in the signing procedure, as demonstrated in Table \ref{Table:checkref}. The idea of first evaluating the infinite norm of the vector with higher rejection probability is applicable to signature schemes that use rejection sampling.
%\subsection{PSPM with Early Evaluation}

The parallel algorithm presented in \cite{DBLP:conf/acsac/ZhengHSXZ22} poses difficulties for \textit{early-evaluation}  as it calculates the entire polynomial vector multiplication results simultaneously. To overcome this issue, we introduce a PSPM algorithm in this section that incorporates early evaluation.  Our algorithm includes the computation of $c \cdot \mathbf{s}+\mathbf{y}$ and $\textbf{LowBits}(\mathbf{w} - c \cdot \mathbf{e},2\gamma_2)$ in the evaluating process, enabling us to promptly perform reject checks for each coefficient. If the reject checks fail, the computation is terminated. This approach results in faster signature speeds. Additionally, there are various PSPM algorithms available. In Dilithium3/5, the coefficients of $\mathbf{s}$ and vector $\mathbf{e}$ are stored in separate precomputed tables, allowing for independent early checks of $\mathbf{z}$ and $\mathbf{r}_0$. In contrast, Dilithium2 stores the coefficients of $\mathbf{s}$ and $\mathbf{e}$ in the same precomputed table. Consequently, the early checks for $\mathbf{z}$ and $\mathbf{r}_0$ are performed simultaneously, as depicted in Algorithm \ref{algo-checkr0z}. It is important to note that during rejection checks, verifying the  $\mathbf{r}_0$ always takes precedence over  checking vector $\mathbf{z}$ for all three parameter sets of Dilithium, as previously analyzed.

\input{repetition}
 \input{checkperformanceref}
\input{PSPMref}
\input{checkr0z}
%%% Tailored reduction %%%%%
\section{Tailored Reduction}
\label{sec:Tailored reduction}
We present an optimized  modular reduction tailored  for Dilithium modulus $q = 8380417$, which might be of independent interest and can be applied to optimize the implementations of Dilithium in other platforms.
 % And analyze the reduced value under different conditions.
The modulus $q$ can be represented as $ 2^{23} - 2^{13} + 1$. We can apply a fast specialized reduction algorithm for modulus prime having such a form. We exemplify with the Dilithium prime and the process is shown in Algorithm \ref{algo-specialreduce2}. 
%This Tailored reduction for $\texttt{NTT}(\mathbf{y})$ can be adopted to speed up both software and hardware implementation of Dilithium2.

\input{tailored_reduction2.tex}
\newtheorem{propsition}{Proposition}
\begin{propsition}
If $-2^{40} < z \leq 2^{40}$, then Algorithm \ref{algo-specialreduce2} computes an integer $r$ congruent to a modulo $ q= 2^{23} - 2^{13} + 1$ such that $-2^{31} < r \leq 2^{31}$.
\begin{proof}
If $-2^{40}<z\leq 2^{40}$, in line 1, $p_{1}=\lfloor {z}/{2^{23}} \rfloor < 2^{17}$,
let $ r_{1}= z-2^{23}p_{1} < 2^{23}$, $r=z-qp_{1}=z-(2^{23}-2^{13}+1)p_{1}=(2^{13}-1)p_{1}+r_{1}$,
so
$$\lvert r \rvert \leq \lvert \left(2^{13}-1\right)p_{1} \rvert +  \lvert r_{1} \rvert  \leq (2^{13}-1)2^{17} + 2^{23} < 2^{31}.$$
\end{proof}

\end{propsition}
\input{montreduction.tex}
\subsection{Comparisons}
Montgomery reduction is an efficient algorithm to reduce product in NTT by computing Hensel remainder. The disadvantage of Montgomery reduction is the Hensel remainder $r'$ is congruent to  $z\cdot2^{-32}\bmod q$ instead of representative of the residue class of $z$ modulo $q$. Algorithm \ref{algo-montreduction} presents the pseudocode of Signed Montgomery reduction. This operation involves two bit-shiftings, two multiplications, and one subtraction. In contrast, our Tailored reduction algorithm is more efficient as it only requires one bit-shifting, one subtraction, and one multiplication. This makes it a better choice than Montgomery reduction when dealing with products smaller than $2^{40}$ for NTT with lazy reduction. Furthermore, the Tailored reduction can be implemented with the new AVX-512IFMA instruction in just two instructions, resulting in a lower latency during reduction (see Subsection \ref{subsec:vec NTT} for a detailed discussion).

%%% Implementation Details %%%%%
\section{Implementation Details}
\label{sec:implem}

We present an optimized vectorization implementation of Dilithium for CPUs that both support the AVX2 and AVX-512 instruction sets. In this section, we will thoroughly explore the implementation details of each optimized module.
\input{percentage_of_used_functions_in_keygen_sign_and_verify}

\subsection{Dilithium Software Performance Profiling}
\label{subsec-profiling}

A critical step in software optimization is to identify the performance bottlenecks of the algorithm.  In this section, we utilize the Linux performance analysis tool perf to profile the Dilithium C \texttt{REF} implementation of Dilithium3 parameter set.
The performance data was collected by executing the Dilithium3 codes 1000 times and calculating the average execution time. Table \ref{Table:functionsused} depicts the detailed percentages. \textsf{KeccakF1600\_StatePermute} which is predominantly used in hash functions, is the most time-consuming function in key generation, signing, and verification. This is followed by Montgomery reduction and \textsf{poly\_uniform} and \textsf{poly\_uniform\_eta}, and then $\texttt{NTT}$ and $\texttt{NTT}^{-1}$.  The functions of  \textsf{poly\_uniform} and \textsf{poly\_uniform\_eta} are used to sample coefficients using the rejection sampling method, while the functions of $\texttt{NTT}$, $\texttt{NTT}^{-1}$ and Montgomery reduction are  used for polynomial multiplication. Consequently, we can identify the computation bottleneck functions as polynomial multiplication, hash function, and rejection sampling. In the following sections, we propose a series of optimization techniques for these functions.

\subsection{Data Alignment}
We represent each polynomial as an array of 256 32-bit signed integers. For this representation, we can use the AVX-512 SIMD instruction to vectorize different functions. Alternatively, we can represent this array as an array of 16 512-bit vectors of type \texttt{\_\_m512i} in AVX-512 intrinsics, where the symbol “i” represents integers. In AVX-512 assembly, we store the 256 coefficients in 16 \texttt{zmm} vector registers.
 The 512-bit Intel AVX-512 registers have an alignment requirement of 64 bytes to ensure optimal vectorization. Optimal memory access is achieved when the data starts at an address on a 64-byte boundary, which means that the address in memory is divisible by 64. Therefore, we align all arrays to 64 bytes in our implementation.

\subsection{Vectorization of NTT with AVX-512}
\label{subsec:vec NTT}
We now give details about our AVX-512 parallel implementation of  NTT for Dilithium polynomial ring $\mathbb{Z}_q[x]/(x^{n}+1)$, where $n = 256, q = 8380417$, the modulus $q$  is a 32-bit prime. The whole NTT-based polynomial multiplication is divided into three parts, $\texttt{NTT}$, $\texttt{NTT}^{-1}$, and point-wise multiplication.
\paragraph{Register allocation}Here we introduce our register arrangement. Note that AVX-512 has 32 512-bit vector \texttt{zmm} registers (\texttt{zmm0}-\texttt{zmm31}). If a 32-bit integer is directly stored in a \texttt{zmm} vector register without zero-padded, a \texttt{zmm} register can store 16 32-bit coefficients,  and hence 16 vector registers are enough to load all 256 coefficients. In doing so, we merge the eight levels without reloading coefficients.  Later in the implementation of the butterfly implementation, we will carefully explain why there is no need to reserve 64-bit space for intermediate products. We arrange \texttt{zmm1}-\texttt{zmm16} to store all the polynomial coefficients consecutively. We use \texttt{zmm17} to store the precomputed results $\zeta q^{-1}  \mod 2^{32}$,  and \texttt{zmm18} to store $\zeta$ ($\zeta$ is the twiddle factor). The \texttt{zmm19}, \texttt{zmm20}, and \texttt{zmm21} are used to store temporary computation values.

\begin{figure}[htbp]
	\centering
	\includegraphics[width=1\linewidth]{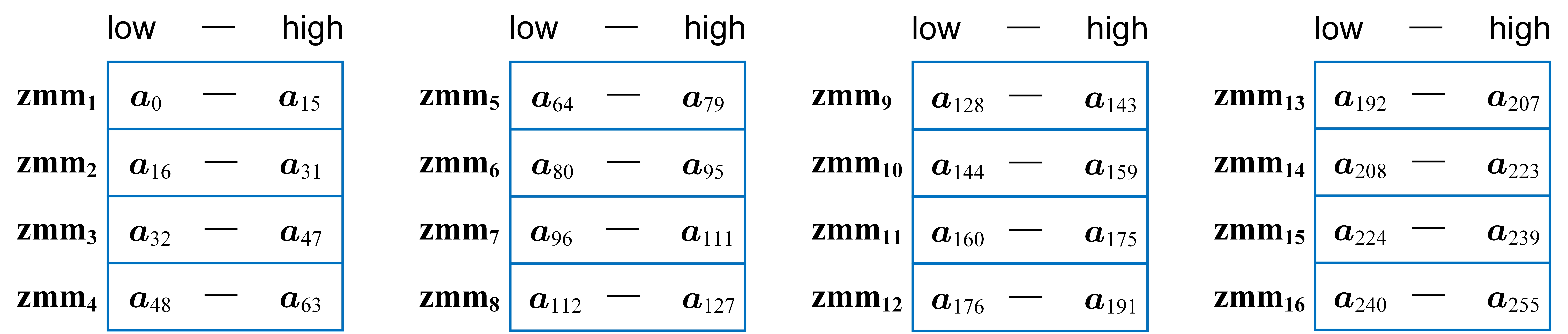}
	
	\caption{The storage coefficients order in \texttt{zmm} registers}
	\label{fig-coeffsorder}
	
\end{figure}
\begin{figure}[htbp]
    \centering
    \subfigure[shuffle eight and four coefficients]{
        \includegraphics[width=2.0in]{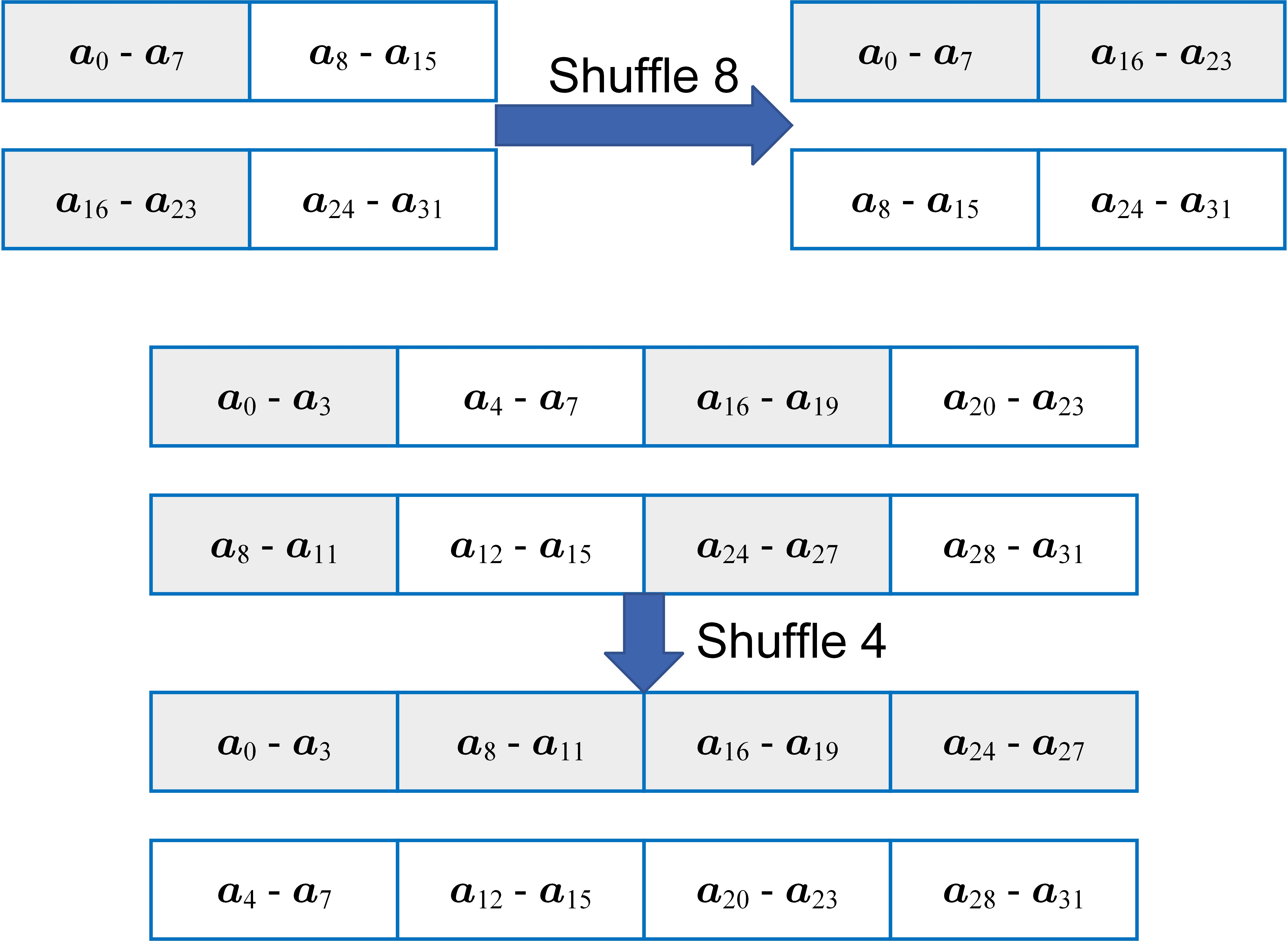}
    }
    \subfigure[shuffle two coefficients]{
	\includegraphics[width=2.0in]{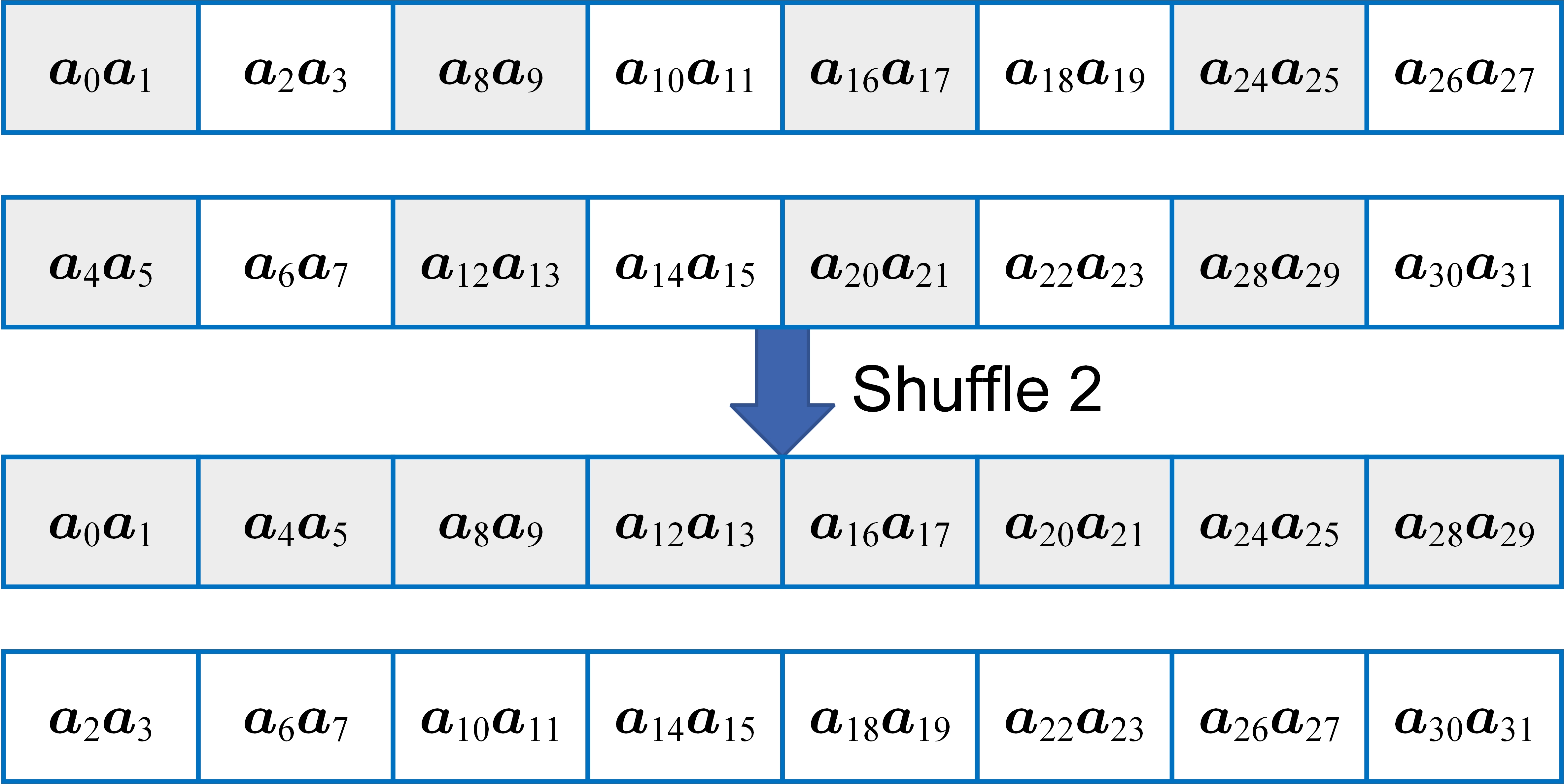}
    }
     \subfigure[shuffle one coefficients]{
	\includegraphics[width=2.8in]{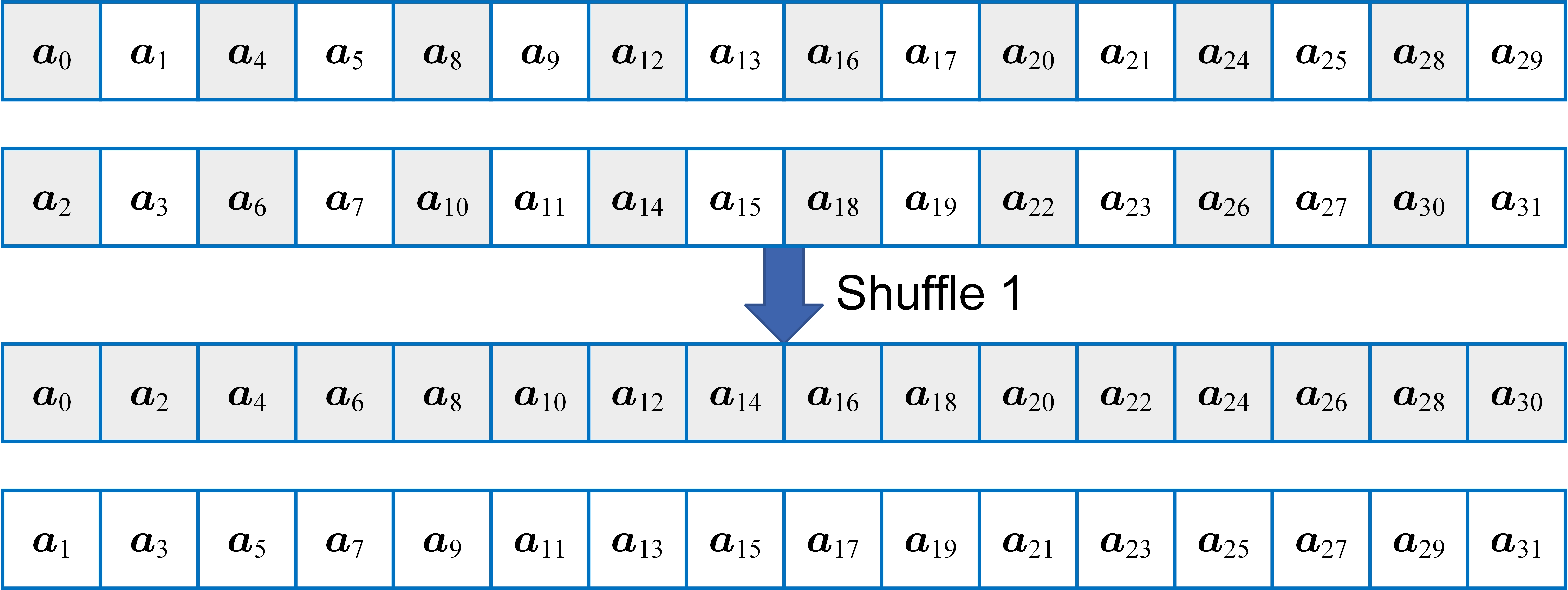}
    }
    \caption{Coefficients shuffling in two vector registers}
    \label{fig-shuffle}
\end{figure}

\paragraph{Coefficients loading and shuffling} We exemplify a polynomial $a_0 + a_1x + ...+a_{255}x^{255}$ as input of \texttt{NTT}. Before the first level, we load the consecutive 16 coefficients in every \texttt{zmm} register as shown in Figure \ref{fig-coeffsorder}. In the first level, the distance of CT butterfly is 128. So the two vector registers \texttt{zmm1} and \texttt{zmm9} perform a pair of butterfly operations, and \texttt{zmm2} and \texttt{zmm10} perform a pair of butterfly operations;  that is, the subscript distance of \texttt{zmm} register is 8. In the second level, the distance is 64. The corresponding registers subscript distance becomes 4. Analogously, the registers subscript distances in the third level and fourth level are two and one respectively. Starting from the fifth level, the distance is 8  while a consecutive 16 coefficients reside in a \texttt{zmm} register. Therefore, in the fifth level, we need to swap the upper 8 coefficients of one register with the lower 8 coefficients of another register. After the fifth level, coefficients are stored in a permutated order in registers. In the sixth level, the distance is 4. The upper four coefficients and the lower four coefficients in every 256-bit data lane are shuffled. Similarly, two coefficients are swapped in every 128-bit data lane in the seventh level and one coefficient is shuffled in every 64-bit data lane in the eighth level. The shuffling process is illustrated in Figure \ref{fig-shuffle}(a), Figure \ref{fig-shuffle}(b), and Figure \ref{fig-shuffle}(c). \textsf{Shuffle8} means to shuffle 8 coefficients, \textsf{Shuffle4} means to shuffle 4 coefficients, \textsf{Shuffle2} means to shuffle 2 coefficients, and \textsf{Shuffle1} means to shuffle one coefficient. We implement \textsf{Shuffle8} using the \texttt{vshufi32x4} instruction. The function of this instruction is to rearrange each 128-bit data lane of the two vector registers $\textsf{a}$ and $\textsf{b}$ through an 8-bit immediate value. We want to rearrange eight consecutive coefficients, which correspond to a 128-bit data lane. According to the instruction pseudocode \footnote{\href{https://www.intel.com/content/www/us/en/docs/intrinsics-guide/index.html}{https://www.intel.com/content/www/us/en/docs/intrinsics-guide/index.html}}, we set the immediate value to \texttt{0x44} and \texttt{0xEE}.

The shuffling of the four coefficients is more complicated because at this time the four consecutive coefficients correspond to a 64-bit data lane. Here we use two permute instructions, one is \texttt{vpermq} and the other is \texttt{vpblendmd}. First, we splice the lower 64-bit in register $\textsf{a}$ and the lower 64-bit in register \textsf{b} using \texttt{vpblendmd}. However, this instruction can only be spliced by the value of the mask register according to the index. Specifically, if we use the \texttt{vpblendmd} directly, the order of the coefficients we will obtain is $\left\{a_0,a_1,a_2,a_3,b_4,b_5,b_6,b_7,a_8,a_{9},a_{10},a_{11},b_{12},b_{13},b_{14},b_{15}\right\}$. This is not the order we want. Therefore, we duplicate the lower 64-bit to the upper 64-bit of register $\textsf{b}$ in every 128-bit data lane and duplicate the upper 64-bit to the lower 64-bit of register $\textsf{a}$ in every 128-bit data lane. We implement this by using \texttt{vpermq} with constant argument \texttt{0x4E} and then using  the \texttt{vpblendmd} instruction to splice the 64-bit data lane in the two registers through the mask register. Here,  we use the \texttt{kmovw} instruction to store \texttt{0x0F0F} into mask register \texttt{k6}. For the permutation of two coefficients, we use \texttt{vpunpcklqdq} and \texttt{vpunpckhqdq}. For the shuffling of one coefficient, because there is no ready-made instruction that can be realized, we adopt the same idea as shuffling four coefficients. First, the upper 32 bits of every 64 bits data lane in register $\textsf{b}$ are obtained by shifting 32 bits to the left. Then use the \texttt{vpblendmd} to splice 32 bits of the two registers with mask register value \texttt{0xAAAA}. For copying the upper 32-bit to the lower 32-bit, we directly use the \texttt{vmovshdup} to copy the upper 32-bit. 
\paragraph{Butterflies} In Section \ref{subsec-NTT}, we introduce NTT and CT/GS butterflies. In the CT butterfly transform, half of the coefficients need to be multiplied by the twiddle factors. Note that the twiddle factors are fixed constants, so we precompute their values and store them in a look-up table. As mentioned earlier, to save multiplication in Montgomery reduction, we also precompute $\zeta q^{-1}  \mod 2^{32}$ and store them in the look-up table. Here we would like to explain why it is not necessary to reserve 64-bit for multiplication results. At the start, the 16 consecutive coefficients are loaded into a \texttt{zmm} register. During the butterfly operation calculation, we split the coefficients that need to be multiplied by the twiddle factor into two parts according to the odd and even subscripts. The odd and even subscript coefficients are stored in two \texttt{zmm} registers. The odd/even coefficient splitting is achieved by copying the upper 32 bits using \texttt{vmovshdup} instruction. After splitting, a register only stores eight coefficients, and each coefficient occupies 64 bits of space. Thus, there is no need to reserve 64 bits of space when loading. Finally, it is reduced to 32-bit by Montgomery reduction. Then the odd-index and even-index coefficients are spliced into a 512-bit vector register by \texttt{vpblendmd} instruction. In this way, although the splitting operation takes some clock cycles, it ensures the maximum degree of parallelism. Generally speaking, this implementation idea is faster than the idea of loading zero-padded 64-bit integers in \cite{lei2023faster}.
\input{tailored_reduction3instr}
 \input{tailoredreductionAVX-512}
 \paragraph{Vectorized Tailored reduction}  We present a vectorized Tailored reduction implementation using AVX-512IFMA instruction. We use this vectorized Tailored reduction implementation in \texttt{NTT}$(\mathbf{t}_0)$ and \texttt{NTT}$(\mathbf{t}_1)$. Previous work implements a four-instruction Montgomery reduction that is both suited for AVX2 and AVX-512 vectorized implementation. The total latency of these four instructions is 12 cycles. In this work, we present a 2-instruction Tailored reduction using AVX-512IFMA \texttt{vpmadd52luq} instruction that can reduce both latency and instruction count and shown in Algorithm \ref{algo-tailord}. This vectorized Tailored reduction reduces the cycle counts down to 6 cycles by eliminating one \texttt{vpmuldq} and one \texttt{vpsubq}.

 \input{MontgomeryreductionAVX-512}
\paragraph{Lazy reduction}Dilithium involves NTT operations on polynomials with small coefficients. We observe that, for CT butterfly of \texttt{NTT} with small coefficients such as $c$ and the noise vectors $\textbf{s}$ and $\textbf{e}$,  the first level does not need to perform Montgomery reduction, because the upper bound data width of $\textbf{s}/\textbf{e}$ is 4 bits, and the multiplication of a 4-bit coefficient and a 23-bit twiddle factor will not exceed 32 bits. $c$ is a small polynomial with only $\tau$ $\pm1$, so the product of a 1-bit coefficient and  a 23-bit twiddle factor will not exceed 32 bits as well. Specifically, we do not need to perform modular reductions in the first level of $\texttt{NTT}(c)$, $\texttt{NTT}(\textbf{s})$ and $\texttt{NTT}(\textbf{e})$. For $\texttt{NTT}(\textbf{t}_0)$ and  $\texttt{NTT}(\textbf{t}_1)$ in all the three security levels of Dilithium2/3/5, as well as  $\texttt{NTT}(\textbf{y})$ in Dilithium2, in the first level of $\texttt{NTT}$  we only need to perform the above tailored  reduction algorithm instead of Montgomery reduction. For instance, in Dilithium2, where $\gamma_{1} = 2^{17}$, the data width of vector $\textbf{y}$ is 18-bit. The product of vector $\textbf{y}$ and the twiddle factor multiplication is a 41-bit integer in $(-2^{40}, 2^{40}]$. Hence,  we use the Tailored reduction Algorithm \ref{algo-specialreduce2}  proposed above. Specifically, in this case,   we do not need to completely reduce the coefficient to $\mathbb{Z}_q$ in the first level of \texttt{NTT}, our only requirement is to prevent the coefficient from overflowing. Starting from  the second level, the product will be reduced by Montgomery reduction.

\subsection{Hashing}
Dilithium makes use of XOF to expand seeds and sample polynomials. \texttt{SHAKE-128} is used to generate matrix $\textbf{A}$, and \texttt{SHAKE-256} is used to generate vectors $\textbf{s}, \textbf{e}$ and $\textbf{y}$. As we discussed in Section \ref{subsec-profiling}, hashing is an expensive operation in the entire scheme. The previous AVX2 implementation used a 4-way  \texttt{SHAKE-128} and  \texttt{SHAKE-256};  that is, they use a vectorized \texttt{SHAKE} implementation that operates on 4 parallel sponges and hence can absorb and squeeze blocks in and out of these 4 sponges at the same time \cite{avanzi2022dilithium}.
We use the AVX-512 implementation and can calculate and generate 8 hash results at the same time due to the expansion of the register bit width.  % We used the SPHINCS+ AVX-512 open source code \footnote{\href{https://github.com/DorAlter/sphincsplus/tree/AVX-512-implementation}{https://github.com/DorAlter/sphincsplus/tree/AVX-512-implementation}}.
 %There already exists highly-optimized 8-way AVX-512 implementations of \texttt{SHAKE-128} and \texttt{SHAKE-256} \cite{alter2021optimizing}.
We embedded this 8-way hash implementation into the expansion of matrix $\textbf{A}$, vector $\textbf{y}$, and vector $\textbf{s}, \textbf{e}$. Dilithium uses \texttt{SHAKE-256} to generate arbitrary length random bytes which is the function $H$. We implement the \texttt{SHAKE-256} using AVX-512. We use five \texttt{zmm} registers to store the 1600-bit keccak state. Each register stores five 64-bit states in its five 64-bit data lanes, while the remaining three data lanes are zero. In this way, we can achieve 5-way parallelism compared with sequential implementation using C.

\subsection{Parallel Rejection Sampling}
The rejection sampling process generates a 23-bit random number by sampling and then checks whether it is greater than or less than $q$ using conditional judgment. If the number is greater than  $q$, it is rejected,  and if it is less than $q$, it is accepted. To obtain the 23-bit random number, the byte stream obtained by hashing needs to be spliced, and then the random number is accepted or rejected sequentially. This process poses a challenge to vectorizing rejection sampling. The previous method used by AVX2 was to create a two-dimensional array of size $2^8\times8=2048$, which stored all possible acceptance positions for 8 32-bit integers in a 256-bit vector register. However, this method is not suitable for AVX-512 implementation because a vector register in AVX-512 can store 16 32-bit integers, requiring a two-dimensional array of size $2 ^{16}\times16=1048576$, which is not feasible for AVX-512 implementation. Therefore, a more space-efficient implementation method was used.% based on the idea in  \cite{gueron2016speeding}.

\begin{figure}[htbp]
\centerline{\includegraphics[scale=.30]{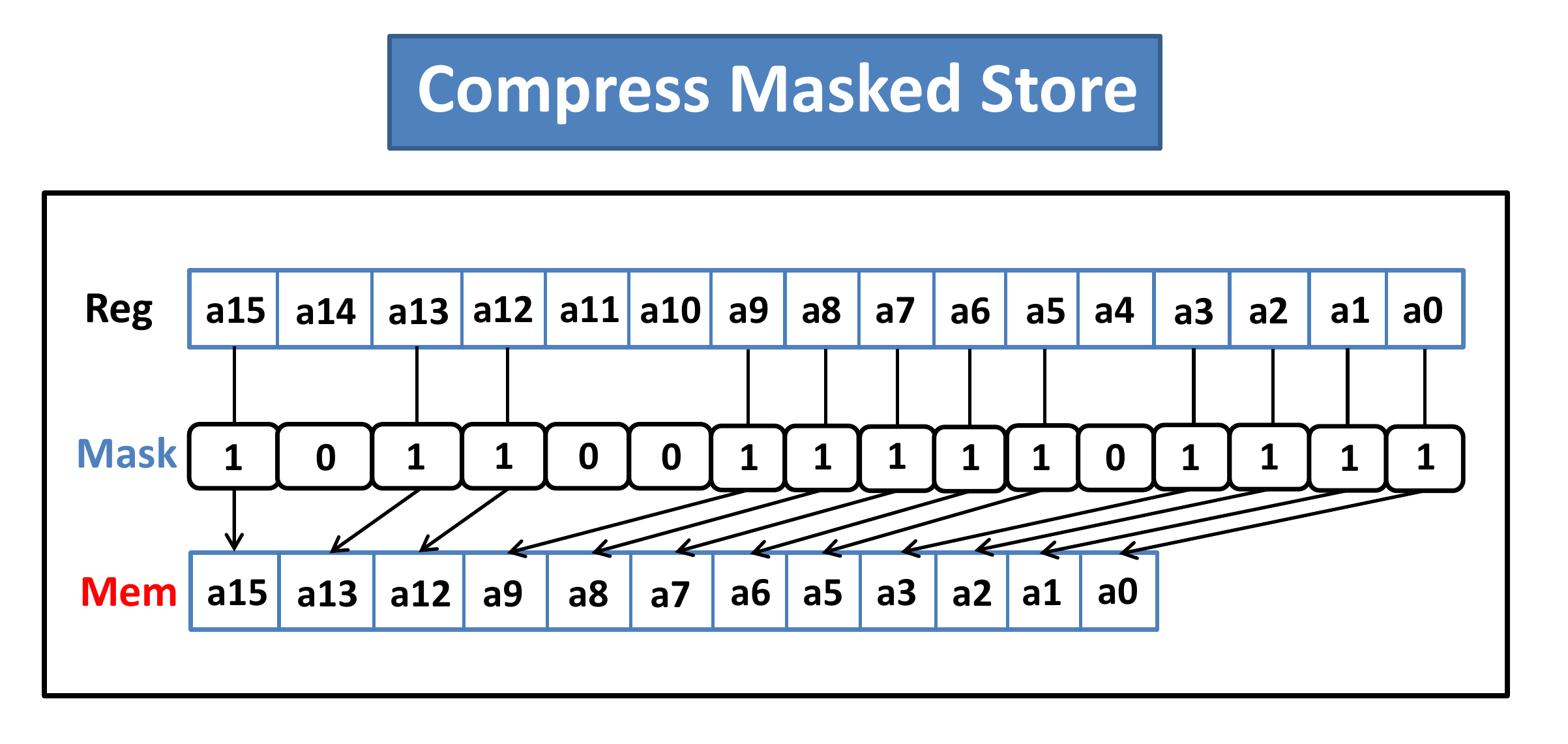}}
\caption{The \texttt{\_mm512\_mask\_compressstoreu\_epi32} function.}
\label{fig-compress}
\end{figure}

 One of the main concepts of rejection sampling is to compare numbers in all positions with $q$ and then store them in order. Fortunately, AVX-512 has a built-in function called \texttt{\_mm512\_mask\_compressstoreu\_epi32}, which stores 32-bit integers in their corresponding positions sequentially through the values of the mask register. This allows us to compressively store the values and meet our requirements. The function is described in Figure \ref{fig-compress}. We can also set the mask register using the function \texttt{\_mm512\_cmp\_epi32\_mask}. By setting the comparison operand value of the  \texttt{\_mm512\_cmp\_epi32\_mask} function to \textsf{\_MM\_CMPINT\_LT}, we compare the values of the input vector register $\textsf{a}$ and vector register $\textsf{b}$. If $\textsf{a}$ is smaller than $\textsf{b}$, we set the value of the mask register at the corresponding position to 1, otherwise, we set it to 0. Note that the mask register is a 16-bit binary integer. We can determine how many coefficients are received in a vector register by counting the number of 1's in the mask register using the function \texttt{\_mm\_popcnt\_u32}.

\begin{figure}[htbp]
    \centering
    \subfigure[\textsf{\_mm512\_mask\_blend\_epi64}]{
        \includegraphics[width=2.2in]{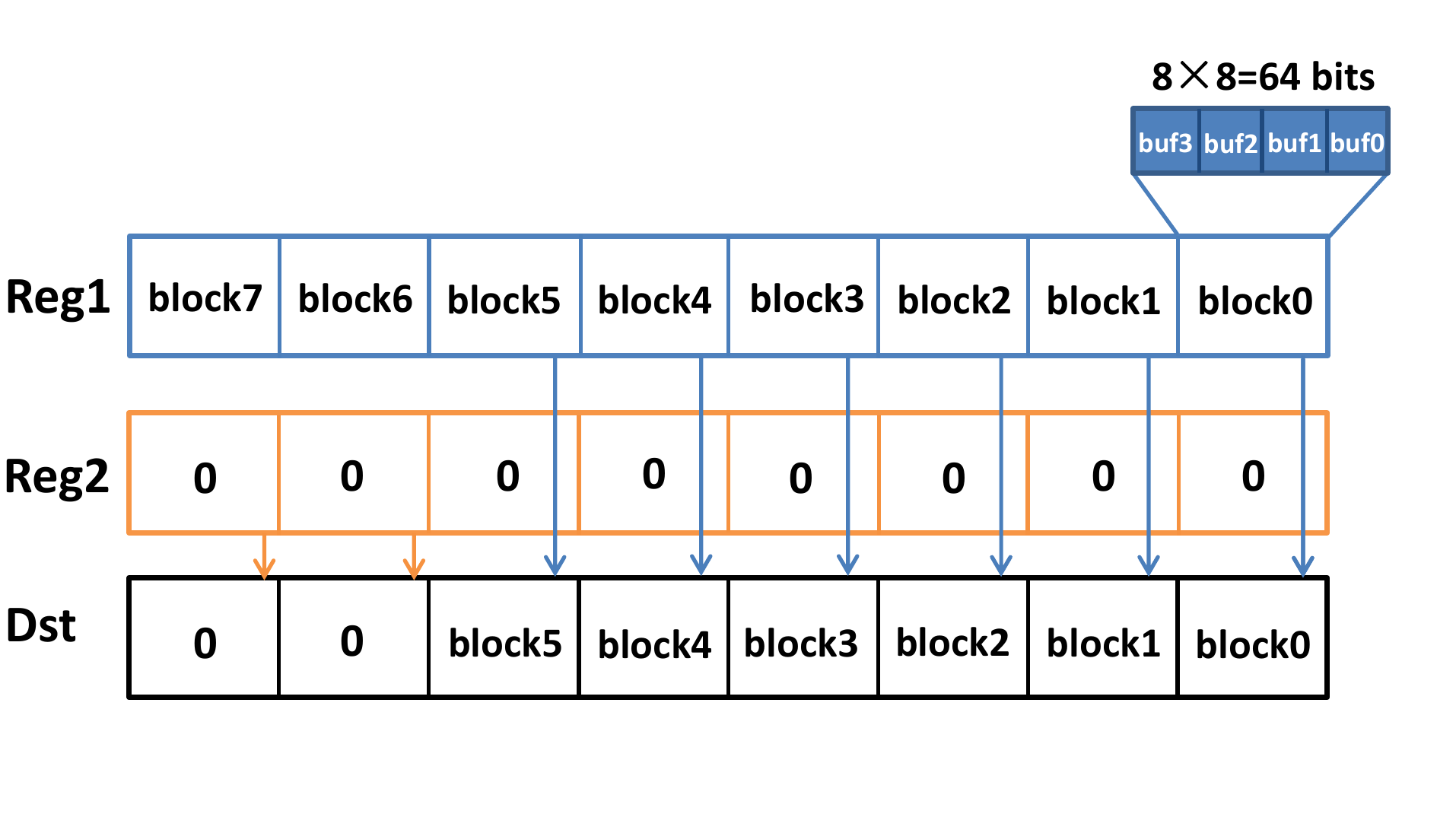}
    }
    \subfigure[\textsf{\_mm512\_permutexvar\_epi8}]{
	\includegraphics[width=2.2in]{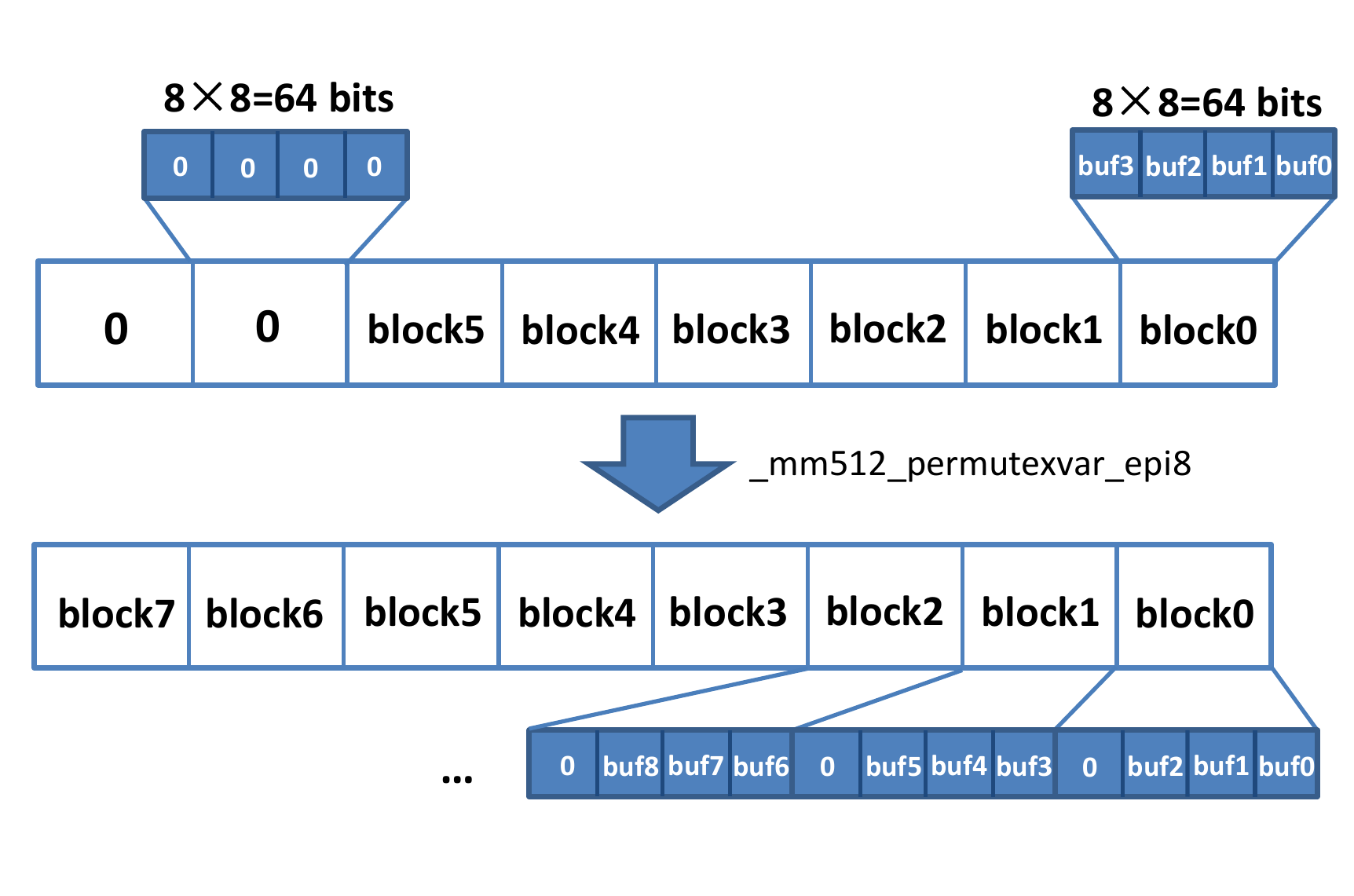}
    }
    \caption{Packing random byte stream}
    \label{fig:rejbuf}
\end{figure}
We optimized the vectorized implementation of generating 23-bit random integers to reduce the number of calls to \texttt{SHAKE-128}. Since we only need 48 out of the 64 bytes streams loaded to obtain 16 23-bit numbers, we should avoid wasting the extra 16 bytes generated by SHAKE-128.

To achieve this, we first initialize a vector register with all zeros, and then use the functions \texttt{\_mm512\_permutexvar\_epi8} and \texttt{\_mm512\_mask\_blend\_epi64} to adjust and splice this all-0 register and the register loaded with 64-byte random byte streams. We illustrate this process in Figure \ref{fig:rejbuf}(a) and Figure \ref{fig:rejbuf}(b). The upper $6\times64=384$ bits are the random byte streams, and the lower $2\times64=128$ bits are zeros. By adjusting the order of the spliced vector registers in the 8-bit data lane using the function \texttt{\_mm512\_permutexvar\_epi8}, we can obtain three consecutive random bytes of every four bytes, and the last byte of the four bytes is just 0. Then, we use \texttt{\_mm512\_and\_si512} to perform a bitwise AND with 23 ones to obtain 16 23-bit random integers.

The above describes the rejection sampling process for generating numbers in the range $[0,q)$. However, in Dilithium, there is also rejection sampling of numbers in the range $[-\eta,\eta]$. We have also optimized the previous AVX2 implementation for this purpose. In our implementation, we first separate the high 4 bits and low 4 bits of each 8-bit random byte, and then use the \texttt{\_mm512\_cmp\_epi32\_mask}  function to judge and store the high 4 bits and low 4 bits separately using mask registers. To ensure the correctness of the test vector, we also adjust the order of the high 4 bits and low 4 bits accordingly. 

\subsection{Expanding Matrix $\textbf{A}$ and Sampling Vectors}
 We present an 8-way \textsf{poly\_uniform\_8x} function to sample 8 polynomials in $R_q$ simultaneously,  using 8-way \texttt{SHAKE-128} and parallel rejection sampling. For the expansion of matrix $\mathbf{A}$, in Dilithium2 where $k=l=4$, we can directly call the \textsf{poly\_uniform\_8x} function twice to generate 4 row vectors. In Dilithium3, where $k=6,l=5$, \textsf{poly\_uniform\_8x} is called four times to generate 30 polynomials of 6 row vectors. In Dilithium5, \textsf{poly\_uniform\_8x} is called eight times to generate 56 polynomials of 8 row vectors. Similarly, for sampling vectors, we propose an 8-way function \textsf{poly\_uniform\_eta\_8x} and \textsf{poly\_uniform\_gamma1\_8x} using 8-way \texttt{SHAKE-256} to sample vectors $\textbf{s}/\textbf{e}$
and $\textbf{y}$ respectively.  

\subsection{Implementing PSPM-TEE }
%%%这一块要重新写，增加early check的实现思路,记得增加表格（测试表格？）
This work implements AVX2 and AVX-512 for PSPM-TEE. In original PSPM implementation from \cite{DBLP:conf/acsac/ZhengHSXZ22}, coefficients were packed into 64-bit words. However, to ensure consistency in the data lane of the vector register and make it easier to operate on the same size operand, we chose to pack coefficients into 32-bit words. This eliminates the need to zero-extend 32-bit coefficients to 64-bit and simplifies the vectorization of PSPM implementation.  For Dilithium2/3/5, we provide a specific description of the implementation of the parallel small polynomial algorithm for Dilithium3 parameters, where $k=6, l=5$. Our implementation is based on the parallel small polynomial parameter sets shown in Table \ref{tab-parapram}.
\input{preandevaluate}

Firstly, we introduce the splicing of the noise vector $\textbf{s}, \textbf{e}$. Although each coefficient of $\textbf{s}$
and $\textbf{e}$  lies in the range of $[-4,4]$, the coefficients grow by $2\tau U$ after the addition operation in Algorithm \ref{algo-parallel}, where $U=4, \tau=49$ in Dilithium3. As a result, the upper bound of coefficients is $392$. Therefore, each coefficient needs to set aside at least 9 bits for storage. One 32-bit word can pack up to 3 polynomial coefficients. Therefore, vectors $\textbf{s}, \textbf{e}$ need two precomputed tables to store all coefficients.

 The preparing process is implemented using intrinsic functions because it is easily vectorizable. However, the loop operation in Algorithm \ref{algo-parallel} is not suitable for parallel implementation. Therefore, our AVX-512 implementation uses parallel computing to implement the accumulation process through AVX-512 assembly. Specifically, when determining whether challenge polynomial $c$ is 1 or -1, we pass the corresponding array address to AVX-512 assembly and perform parallel addition.  Combining with the parallelism achieved by Algorithm \ref{algo-checkr0z}, the implementation of $c\textbf{s}$ can achieve a maximum of $8\times3=24$ parallelism at most.

We implemented the evaluating process of extracting computation results from the 32-bit packed words using the intrinsic functions. To perform the conditional check of vector coefficients, we used the \texttt{\_mm512\_cmp\_epi32\_mask} function, which allows us to check 8 coefficients in parallel and obtain a 16-bit mask for every 32-bit data lane. If the mask is non-zero, the function immediately returns 1.

\subsection{Vectorized Packing}
\paragraph{Obstacle in vectorizing packing}
In Dilithium implementation, polynomial vectors need to be encoded as byte strings (packing) and vice versa (unpacking). We have completed the vectorization of unpacking of $\mathbf{z}$ and packing of $\mathbf{w}_1$ using AVX-512. To ensure that our optimized implementation works on all platforms and matches the NIST  Known Answer Tests (KAT) test vectors, we faced a difficulty in vectorizing polynomial packing and unpacking. Directly vectorizing the packing/unpacking process is not feasible. For instance, a 512-bit vector register can store 16 coefficients, and bit-wise instructions are operated on two vector registers. If register $\textsf{r}_1$ stores coefficients $a_0-a_{15}$, $\textsf{r}_2$  stores coefficients $a_{16}-a_{31}$. A pair of coefficients $a_0$ and $a_{16}$ are packed, whereas we need $a_0$ and $a_1$. Therefore, direct vectorization is not possible.
\begin{figure}
  \centering
  \includegraphics[width = 0.5\textwidth]{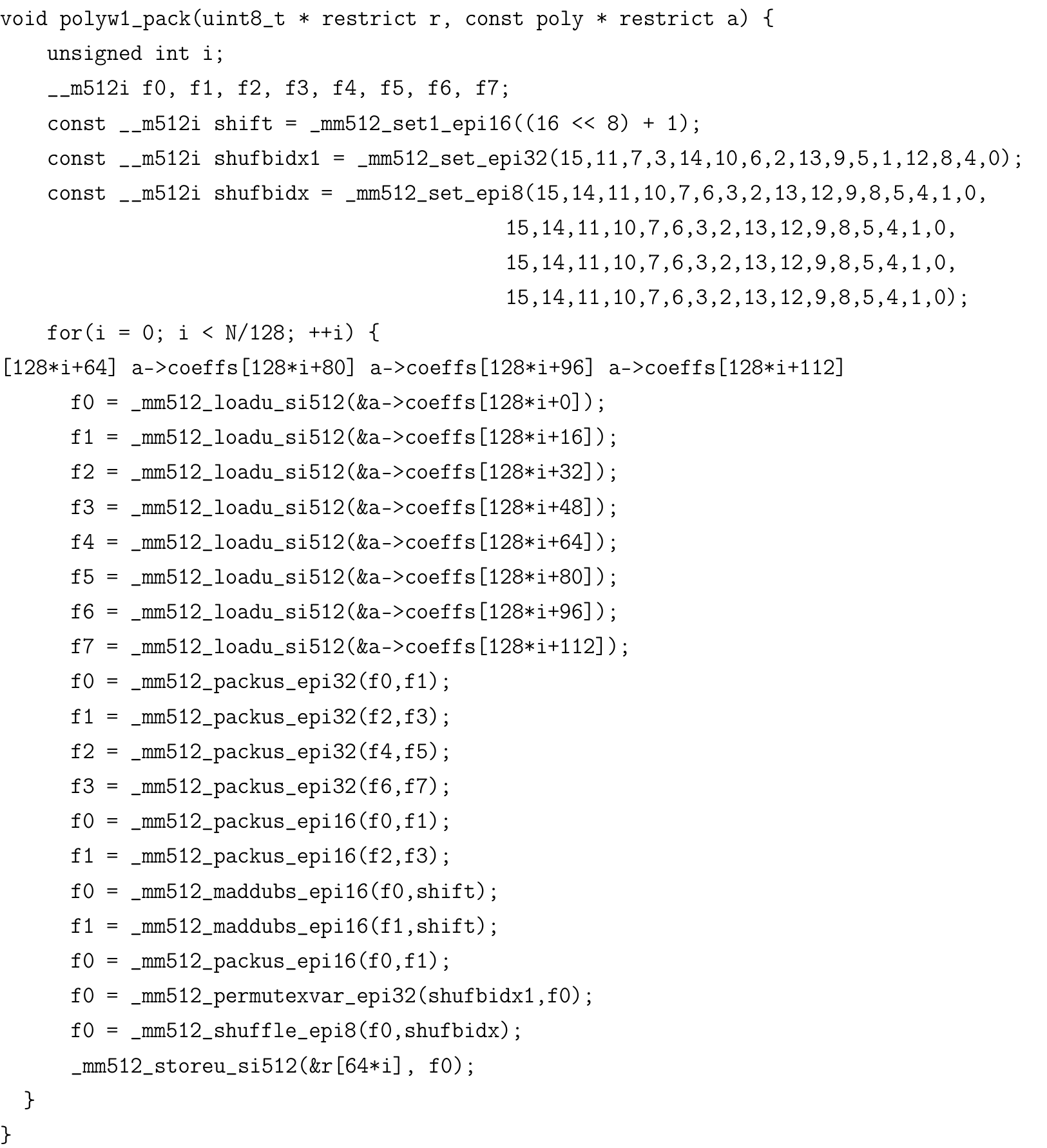}
  \caption{Packing $\mathbf{w}_1$ function using AVX-512.}
  \label{fig:pack_w}
\end{figure}
\paragraph{How to vectorize packing}
The vectorization of unpacking $\mathbf{z}$ using AVX-512 is similar to parallel rejection sampling. For packing of $\mathbf{w}_1$. We take Dilithium3/5 parameter for example, the coefficient range of $\mathbf{w}_1$ is $[0,15]$. Every two $\mathbf{w}_1$ coefficients can be packed into one byte. We need to sequencely pack 4-bit $\mathbf{w}_1$ coefficients in one \texttt{zmm} register and store back to memory in 8-bit data lane. As shown in Figure \ref{fig:pack_w}, we use a series of convert instructions to convert 32-bit coefficients to packed 8-bit coefficients. Since there is no instruction to directly convert 32-bit to 4-bit, we propose to shift of odd indices coefficients to the left by 4 bits and then pack with the even indices 4 bits. Finally, we adjust the order by using permutation instruction to ensure the the correctness of the KAT test.

% A \texttt{zmm} register can store 128 4-bit $\mathbf{w}_1$ coefficients. Therefore, 128 polynomial coefficients are loaded continuously in each loop. We use \texttt{\_mm512\_packus\_epi32} in Figure \ref{fig:pack_w} to convert packed signed 32-bit integers to packed 16-bit integers. Next, we use \texttt{\_mm512\_packus\_epi16} to convert packed 16-bit integers to packed 8-bit integers, which results in 64 coefficients stored in each vector, with each coefficient having a width of 8 bits. We then use \texttt{\_mm512\_maddubs\_epi16} to shift the coefficients with odd indices by 4 bits to the left, so that adjacent coefficients are combined into one byte. Finally, we use \texttt{\_mm512\_maddubs\_epi16} to merge all coefficients into a 512-bit vector. However, after the above operation, we change the order of the coefficients. Therefore, we use the intrinsic functions \texttt{\_mm512\_permutexvar\_epi32} and \texttt{\_mm512\_shuffle\_epi8} to reorder the coefficients to ensure the correctness of the KAT test.

%%%% Experiment Results and Comparison %%%%
\section{Experiment Results and Disscussions}
\label{sec:experi}
We implemented all three security levels of Dilithium,  using both C language and Intel AVX-512 assembly and AVX-512 intrinsic functions. For a more comprehensive comparison, we provide both the Round3 submission version of Dilithium \cite{avanzi2022dilithium} and the FIPS 204 version of Dilithium \cite{dilidraft2023}, known as ML-DSA. We also optimized the previous Round3 submitted AVX2 code using the presented optimization technique. Our optimized vectorization implementation has successfully passed the NIST Known Answer Tests, thereby confirming its compatibility across all platforms. We proceed to conduct a thorough performance evaluation, highlighting the improvements achieved through the optimizations discussed in Section \ref{sec:implem}.
%Because memory is not a constraint in desktop computers, here we only report the results of execution time.
The Round3 Dilithium codes are collected from \href{https://csrc.nist.gov/Projects/post-quantum-cryptography/selected-algorithms-2022}{https://csrc.nist.gov/Projects/post-quantum-cryptography/selected-algorithms-2022}. The FIPS204 ML-DSA codes are collected from \href{https://github.com/pq-crystals/dilithium/tree/standard}{https://github.com/pq-crystals/dilithium/tree/standard}. The compiler is gcc-9.4.0 and the optimization flags are  \textsf{-Wshadow -Wpointer-arith -mavx2 -mAVX-512F -mAVX-512vbmi -mAVX-512bw -mAVX-512cd -mAVX-512vl -mpopcnt -maes -march=native -mtune=native -O3}. The benchmark experiments were conducted on a desktop machine with Ubuntu 20.04 operating system and Intel(R) Core(TM) i7-11700F CPU (Rocket Lake) running at 2.5GHz. As usual, we disable the TurboBoost and Hyper-Threading to ensure the reproduction of the experiments. Each experiment is repeated 100000 times, and we present the median results.
% \subsection{Tailored  Reduction Performance}
% We implemented Tailored reduction using AVX2 and AVX-512 in the first level of \texttt{NTT}. We make a comparison with \texttt{NTT} using Montgomery reduction. To demonstrate the true benefits of Tailored reduction in the signing process, we compared the original AVX2 signing implementation with the AVX2 implementation using Tailored reduction in the \texttt{NTT}$(\mathbf{t}_0)$.  In Table \ref{Table:tailoredifma}, Sign-original refers to the implementation of signing with NTT using Montgomery reduction. Sign-opt1 refers to the implementation of signing with NTT using Tailored reduction by AVX-512F. Sign-opt2 refers to the implementation of signing with NTT using Tailored reduction by AVX-512IFMA. Table \ref{Table:tailoredifma} shows that Tailored reduction provides a speedup of 3\% in signing. Additionally, our implementation of Tailored reduction using AVX-512IFMA is much faster than AVX-512F, confirming that IFMA does indeed improve the performance of Tailored reduction implementation.
% \input{table/Tailoredreductionperf}
\subsection{Polynomial Multiplication Performance}
Table \ref{Table:pmperf} presents performance results for polynomial multiplication within Dilithium. We report benchmark results for polynomial-vector multiplication $c\cdot\mathbf{s}$ and $c\cdot\mathbf{e}$ using NTT and PSPM techniques. On Intel CPUs, PSPM achieves speed improvements ranging from 47\% to 66\% compared to NTT, for both AVX2 and AVX-512 implementations. Furthermore, the data indicates that our AVX-512 implementations of $c\cdot\mathbf{s}$ and $c\cdot\mathbf{e}$ using the NTT technique are 53\% faster than their AVX2 counterparts.

\input{performance_of_PSPM.tex}
\subsection{PSPM-TEE Performance}
Table \ref{Table:pspmperf} illustrates the performance of the Improved PSPM algorithm. When we apply the Improved PSPM, our AVX2 implementation achieves a speedup of approximately 7.7\% for Dilithium3, while Dilithium5 experiences a more modest acceleration of only 2.9\%. These results demonstrate the discernible advantages of the improved PSPM algorithm in enhancing the signing procedure of Dilithium. Consequently, we incorporate the improved PSPM algorithm in the implementation of Dilithium using AVX-512 as well.
\input{PSPMperformance.tex}

\subsection{Other Vectorization Functions Performance}
We conducted an experiment primarily to evaluate the performance of our AVX-512 vectorized functions within the context of Dilithium, as detailed in Table  \ref{Table:vecfunc}. Our benchmark results encompass two versions of \texttt{SHAKE-256}: the parallel version and the sequential version. The parallel version of \texttt{SHAKE-256} generates four or eight hashing results simultaneously, while the sequential version generates only one hashing result. Since we can store five 64-bit states in one AVX-512 register, the sequential version exhibits 5-way parallelism.
\input{vecfunction.tex}

In the case of NTT, we adopted an efficient approach for loading coefficients, enabling us to load 16 coefficients simultaneously, a significant improvement over AVX2's four-coefficient loading capacity. This optimization resulted in a substantial 16-way parallelism in our NTT AVX-512 implementation, effectively reducing memory access. Consequently, we achieved a remarkable acceleration factor of nearly 14 times in $\texttt{NTT}$. Similarly, in $\texttt{NTT}^{-1}$, we realized a commendable speedup of nearly 18 times. The improvements in NTT are primarily attributed to the inherent vectorization capabilities of AVX-512, as well as our well-structured instruction scheduling and efficient utilization of registers, which significantly reduce load and store operations through layer merging technology. In comparison to the NTT AVX-512 implementation in \cite{lei2023faster}, which demonstrated speedups of 12.13x, 13.46x, and 11.50x in $\texttt{NTT}$, $\texttt{NTT}^{-1}$, and polynomial pointwise multiplication respectively, our AVX-512 implementation exhibits superior performance due to its enhanced parallelism.

% In the polynomial pointwise multiplication function, the parallelism is reduced by half during the calculation due to the large number of  $32 bits\times32 bits \Longrightarrow64 bits$ multiplication operations, resulting in only an 11-fold acceleration ratio. We also implemented AVX-512's 64-way vectorization on the packing part, and the corresponding data is shown in Table \ref{Table:vecfunc}  in Appendix \ref{appendix:Tables}. However, since Dilithium's AVX2 code does not have a \textsf{polyz\_unpack}  implementation, the 32-way data is not given here. Our 16-way vectorization achieves a 30-fold speedup. The main bottleneck of the \textsf{polyz\_unpack} acceleration is the coefficient's reordering to ensure the correctness of the test vectors. The \textsf{polyw1\_pack} itself is relatively simple. The C implementation is already fast, so the 16-way implementation does not bring much acceleration.

\subsection{Scheme Performance}

In this work, we pursued peak performance by employing a range of optimization techniques in the implementation of Dilithium. These optimizations encompassed various aspects, including enhancements in NTT, rejection sampling, decomposition, computing hints, bit-packing, and more. Table X provides a summary of cycle counts and comparisons for all three security levels of Round3 Dilithium, encompassing key generation (KeyGen), signing (Sign), and verification (Verify).

\cite{dilidraft2023} presented benchmark results for Round3 Dilithium3; however, due to their unavailability of open-source code, our comparison focused on speedups. In key generation, signing, and verification, we achieved speedups of 65.1\%, 52.8\%, and 56.2\%, respectively, surpassing their speedups of 33.6\%, 43.2\%, and 40.1\%. These performance improvements were primarily driven by our optimized NTT implementation and the introduction of the PSPM-TEE algorithm. Additionally, our sequential SHAKE-256 implementation using AVX-512 contributed to the overall performance enhancements.
\input{schemeperformance.tex}
We enhanced Round3 AVX2 implementation by incorporating the improved PSPM and tailored reduction techniques, resulting in a speedup of 3\% to 8\% in the signature procedure. In our Dilithium AVX-512 implementation, certain parts have not yet been vectorized, such as hash functions other than polynomial sampling. Consequently, the overall improvement in signature speed cannot exceed twice the AVX2 software speed. Nonetheless, our speedup primarily stems from the vectorization of specific functions and the optimization techniques we introduced.
\input{AVX2optimizedperformance}
Dilithium was chosen as one of the digital signature standards on July 22th, 2022. On August 24th, 2023, NIST published the standardization document FIPS 204 \cite{dilidraft2023}, which aligns with the Dilithium scheme. There exist several distinctions between the FIPS 204 ML-DSA Standard and Round3 Dilithium, and we have also implemented the FIPS 204 scheme ML-DSA, providing benchmark results in Section \ref{tab:standscheme}.
\input{standardperformance.tex}

\subsection{Discussions about Side-Channel Security and Memory Cost}
 Constant-time implementation (CTI) was not the focus of this work, but we indeed take it in mind.  We have carefully avoided using branching statements depending on secret information, and we have not used the modulo operator \%. For the side-channel security of the   PSPM-TEE technique, we have the following observations. On the one hand,  as the intermediate hashing $c$'s rejected with the tailored early evaluation are never output, the intermediate values are actually blinded to an outside observer. On the other hand, the PSPM technique consolidates coefficients of identical dimensions from multiple small polynomials into a single word for operations. This approach can potentially introduce greater complexity and obstacles for side-channel attacks when compared to traditional NTT technique. 

  For space cost, our implementation pre-calculates the tables in improved PSPM, which requires an additional 8192 bytes of storage space in Dilithium3/5 and 4096 bytes in Dilithium2. However, our implementation of parallel rejection sampling saves 1048576 bytes. Overall, our implementation significantly reduces the required space compared to the previous AVX2 implementations.

\subsection{Deployment on Other Platforms}
Due to the absence of AVX-512 compatible CPUs, there may be questions surrounding deployment on alternative platforms. However, it is still worthwhile to consider the performance enhancements achievable on x86-64 CPUs utilizing the latest AVX instructions. Moreover, some of the optimization techniques we propose can be implemented in other systems. The upgraded 32-bit version of the PSPM algorithm is especially advantageous for ARM Cortex-M4 implementation, as it solely utilizes 32-bit general registers. Additionally, the Tailored reduction technique exclusively employs subtraction and multiplication, making it readily deployable on other platforms. Furthermore, our implementation addresses issues that arise when vectorizing serial processes or functions, ensuring the accuracy of test vectors.
\section{Conclusion}
\label{sec:conclu}
This paper demonstrates the potential of AVX-512 in accelerating the implementation of Dilithium. Specifically, we enhance PSPM through the introduction of PSPM-TEE, significantly expediting the Dilithium signing process. We illustrate how tailored reduction can be applied to Dilithium's modulus, presenting a fast implementation using AVX-512IFMA. We extensively vectorize numerous functions within Dilithium, with a particular focus on addressing performance bottlenecks such as polynomial multiplication, hashing, and more. In summary, we provide a fully vectorized implementation of Dilithium utilizing AVX-512. Leveraging these optimization techniques, our implementation achieves substantial speed improvements over previous AVX2 implementations, thereby establishing the most efficient Dilithium implementation on the x86-64 platform to date.

\normalem
\bibliography{ref}

%%%% APPENDIX %%%%
\appendix
%\section{PSPM with Early Evaluation pseudocode for Dilithium3/5}
\subsection{PSPM with Early Evaluation pseudocode for Dilithium3/5}
\label{appendix:Codes}
\input{checkz}
\input{checkr0}

\newpage
\newpage
%\section{Tables}
\label{appendix:Tables}

\subsection{Parameter for PSPM}
\input{parallel_parameter.tex}

\newpage

\end{document}

%% file: keygen.tex
\begin{algorithm}[H]
  \caption{ML-DSA.$\mathsf{KeyGen}()$}
  \label{algo-keygen}
  \small
  \begin{algorithmic}[1]
    \Statex \textbf{Input:} $\zeta \leftarrow\{0,1\}^{256}$
    \Statex \textbf{Output:} Public and secret keys $\left(p k=\left(\rho, \mathbf{t}_1\right), s k=\left(\rho, K, t r, \mathbf{s}_1, \mathbf{s}_2, \mathbf{t}_0\right)\right)$
    \State $\left(\rho, \rho^{\prime}, K\right) \in\{0,1\}^{256} \times\{0,1\}^{512} \times\{0,1\}^{256}:=\mathrm{H}(\zeta)$ \Comment H is instantiated as SHAKE-256
    \State $\mathbf{A} \in \mathcal{R}_{q}^{k \times \ell} := \mathsf{ExpandA}(\rho)$ \Comment $\mathbf{A}$ is generated and stored in NTT Representation as $\hat{\mathbf{A}}$
    \State $\left(\mathbf{s}, \mathbf{e}\right) \in S_\eta^{\ell} \times S_\eta^k:=\mathsf{ExpandS}\left(\rho^{\prime}\right)$
    \State $\mathbf{t}:=\mathbf{A} \mathbf{s}+\mathbf{e}$ \Comment Compute $\mathbf{A s}$ as $\texttt{NTT}^{-1}\left(\hat{\mathbf{A}} \circ \texttt{NTT}\left(\mathbf{s}\right)\right)$
    \State $(\mathbf{t}_{1},\mathbf{t}_{0}) := \mathsf{Power2Round}_{q,d}(\mathbf{t})$
    \State $tr \in\{0,1\}^{256}:=\mathrm{H}\left(\rho \| \mathbf{t}_1\right)$
    \State \textbf{return} $\left(p k=\left(\rho, \mathbf{t}_1\right), s k=\left(\rho, K, t r, \mathbf{s}, \mathbf{e}, \mathbf{t}_0\right)\right)$
  \end{algorithmic}
\end{algorithm}

%% file: sign.tex
\begin{algorithm}[H]
  \caption{ML-DSA.$\mathsf{Sign}(sk, M)$}
  \label{algo-sign}
  \small
  \begin{algorithmic}[1]
     \Statex \textbf{Input:} Secret key $s k=\left(\rho, K, t r, \mathbf{s}, \mathbf{e}, \mathbf{t}_0\right)$, Message $M \in \{0,1\}^*$
     \Statex \textbf{Output:} Signature $\sigma=(\tilde{c}, \mathbf{z}, \mathbf{h})$
    
    \State $\mathbf{A} \in \mathcal{R}_{q}^{k \times \ell} := \mathsf{ExpandA}(\rho)$ \Comment $\mathbf{A}$ is generated and stored in NTT Representation as $\hat{\mathbf{A}}$
    \State $\mu \in\{0,1\}^{512}:=\mathrm{H}(tr \| M)$
    \State $\kappa:=0,(\mathbf{z}, \mathbf{h}):=\perp$
    \State$\rho^{\prime} \in\{0,1\}^{512}:=\mathrm{H}(K \| \mu)$ (or $\rho^{\prime} \leftarrow\{0,1\}^{512}$ for randomized signing)
    
    \While {$(\mathbf{z}, \mathbf{h}) = \bot$} \Comment Pre-compute $\hat{\mathbf{s}}:=\texttt{NTT}\left(\mathbf{s}\right), \hat{\mathbf{e}}:=\texttt{NTT}\left(\mathbf{e}\right)$, and $\hat{\mathbf{t}}_0:=\texttt{NTT}\left(\mathbf{t}_0\right)$
        \State $\mathbf{y} \in \tilde{S}_{\gamma_1}^{\ell}:=\mathsf{ExpandMask}\left(\rho^{\prime}, \kappa\right)$
        \State $\mathbf{w} := \mathbf{Ay}$ \Comment $\mathbf{w}:=\operatorname{NTT}^{-1}(\hat{\mathbf{A}} \circ \texttt{NTT}(\mathbf{y}))$
        \State $\mathbf{w}_1 := \mathsf{HighBits}_q(\mathbf{w}, 2\gamma_2)$
        \State $\tilde{c} \in\{0,1\}^{256}:=\mathrm{H}\left(\mu \| \mathbf{w}_1\right)$
        \State $c \in B_\tau:= \mathsf{SamplelnBall}(\tilde{c})$  \Comment{Store $c$ in NTT representation as $\hat{c}=\texttt{NTT}(c)$}
        \State $\mathbf{z} := \mathbf{y} + c\mathbf{s}$ \Comment{Compute $c \mathbf{s}$ as $\texttt{NTT}^{-1}\left(\hat{c} \circ \hat{\mathbf{s}}\right)$}
        \State $\mathbf{r}_0:=\mathsf{LowBits}_q\left(\mathbf{w}-c \mathbf{e}, 2 \gamma_2\right)$ \Comment{Compute $c \mathbf{e}$ as $\texttt{NTT}^{-1}\left(\hat{c} \circ \hat{\mathbf{e}}\right)$}
        \If {$||\mathbf{z}||_\infty \geq \gamma_1 - \beta$ or $||\mathbf{r}_0||_\infty \geq \gamma_2 - \beta$ }
            $(\mathbf{z}, \mathbf{h}) := \bot$
        \Else
            \State $\mathbf{h} := \mathsf{MakeHint}_q(-c\mathbf{t}_0, \mathbf{w} - c\mathbf{e} + c\mathbf{t}_0, 2\gamma_2)$ \Comment{Compute $c \mathbf{t}_0$ as $\texttt{NTT}^{-1}\left(\hat{c} \circ \hat{\mathbf{t}}_0\right)$}
            \If {$||c\mathbf{t}_0||_\infty \geq \gamma_2$ or $\mathsf{NumberOfOne}(\mathbf{h}) > \omega$}
                $(\mathbf{z}, \mathbf{h}) := \bot$
            \EndIf
        \EndIf
        \State $\kappa:=\kappa+\ell$
    \EndWhile

    \State \textbf{return} $\sigma=(\tilde{c}, \mathbf{z}, \mathbf{h})$
  \end{algorithmic}
\end{algorithm}

%% file: verify.tex
\begin{algorithm}[H]
  \caption{ML-DSA.$\mathsf{Verify}(pk, M, \sigma=(\tilde{c}, \mathbf{z}, \mathbf{h}))$}
  \label{algo-verify}
  \small
  \begin{algorithmic}[1]
     \Statex \textbf{Input:} Public key $pk = (\rho, \mathbf{t}_{1})$, Message $M \in \{0,1\}^*$, Signature $\sigma=(\tilde{c}, \mathbf{z}, \mathbf{h})$
     \Statex \textbf{Output:} Result $r \in \{0,1\}$
    
    \State $\mathbf{A} \in \mathcal{R}_{q}^{k \times \ell} := \mathsf{ExpandA}(\rho)$ \Comment{$\mathbf{A}$ is generated and stored in NTT Representation as $\hat{\mathbf{A}}$}
    \State $\mu \in\{0,1\}^{512}:=\mathrm{H}\left(\mathrm{H}\left(\rho \| \mathbf{t}_1\right) \| M\right)$
    \State $c:=\mathsf{SamplelnBall}(\tilde{c})$
    \State $\mathbf{w}'_1 := \mathsf{UseHint}_q(\mathbf{h}, \mathbf{Az} - c\mathbf{t}_1 \cdot 2^d, 2\gamma_2)$ \Comment{Compute as $\texttt{NTT}^{-1}\left(\hat{\mathbf{A}} \circ \texttt{NTT}(\mathbf{z})-\texttt{NTT}(c) \circ \texttt{NTT}\left(\mathbf{t}_1 \cdot 2^d\right)\right)$}
    
    \State \textbf{return} $\tilde{c}=\mathrm{H}\left(\mu \| \mathbf{w}_1^{\prime}\right)$ \textbf{and} $\|\mathbf{z}\|_{\infty}<\gamma_{1}-\beta$ \textbf{and} $\mathsf{NumberOfOne}(\mathbf{h}) \leq \omega$ 
  \end{algorithmic}
\end{algorithm}

%% file: parallel.tex
\begin{algorithm}[H]
  \caption{A parallel index-based polynomial multiplication algorithm with translations}
  \label{algo-parallel}
  \small
  \begin{algorithmic}[1]
    \Statex \textbf{Input:}  $(c, \mathbf{a})$, where $\mathbf{a} = [a^{(0)}, \cdots, a^{(r-1)}]^T \in \mathcal{R}_{q}^r$, every $a^{(j)}=\sum_{i=0}^{n-1} a^{(j)}_{i} \cdot x^{i} \in \mathcal{R}_{q}$, and $c=\sum_{i=0}^{n-1} c_{i} \cdot x^{i} \in B_{\tau}$
    \Statex \textbf{Output:}  $\mathbf{u} = c \cdot \mathbf{a} = [u^{(0)}, \cdots, u^{(r-1)}]^T \in \mathcal{R}_{q}^r$, where $u^{(j)}=c \cdot a^{(j)} = \sum_{i=0}^{n-1} u^{(j)}_{i} \cdot x^{i} \in \mathcal{R}_{q}$
    
    \For {$i \in \{0,1,\cdots, n-1\}$}
      \State $w_i := 0$
      \State $v_i := 0$
      \State $v_{i-n} := 0$
      \For {$j \in (0, 1, \cdots, r-1)$}
        \State $v_{i}:=v_{i} \cdot M+\left(U+a_{i}^{(j)}\right)$
        \State $v_{i-n}:=v_{i-n} \cdot M+\left(U-a_{i}^{(j)}\right)$
      \EndFor
    \EndFor
    \State $\gamma:=2 U \cdot \frac{M^{r}-1}{M-1}$
    \For {$i \in \{0,1,\cdots, n-1\}$}
      \If {$c_i = 1$}
        \For {$j \in \{0,1,\cdots, n-1\}$}
          \State $w_{j} := w_{j} + v_{j-i}$
        \EndFor
      \EndIf
      \If {$c_i = -1$}
        \For {$j \in \{0,1,\cdots, n-1\}$}
          \State $w_{j} := w_{j} + (\gamma - v_{j-i})$
        \EndFor
      \EndIf
    \EndFor
    \For {$i \in \{0,1,\cdots, n-1\}$}
      \State $t := w_i$
      \For {$j \in (0, 1, \cdots, r-1)$}
        \State $u_{i}^{(r-1-j)}:=(t \bmod M)-\tau U(\bmod q)$
        \State $t:=\lfloor t / M\rfloor$
      \EndFor
    \EndFor
    \State \textbf{return} $\mathbf{u} = [u^{(0)}, \cdots, u^{(r-1)}]^T$
  \end{algorithmic}
\end{algorithm}

%% file: repetition.tex
\captionsetup[table]{labelfont={bf}}
\begin{table}[htbp]
\caption{Probability of vector in a good range. }
	\label{Table:rep}
    \centering
	\renewcommand\arraystretch{1}
	\begin{tabular}{*{3}{c}}
		\specialrule{1.0pt}{0pt}{0pt}
		{Scheme}
			& {$\mathrm{Pr}\left(|| \mathbf{z}|| \leq \gamma_1-\beta\right)$}
			& {$\mathrm{Pr}\left(|| \mathbf{r}_0|| \leq \gamma_1-\beta\right)$}
			\\
			\hline
		{Dilithium2}
			& {0.543591}
			& {0.429801}
			\\ \hline
		{Dilithium3}
			& {0.619647}
			& {0.315712}
			\\ \hline
		{Dilithium5}
			& {0.663515}
			& {0.389636}
			\\ \specialrule{1.0pt}{0pt}{0pt}
	\end{tabular}
	
\end{table}

%% file: checkperformanceref.tex
\captionsetup[table]{labelfont={bf}}
\begin{table}[!ht]
    \centering
	\renewcommand\arraystretch{1}
	\caption{Comparative performance of checking $\mathbf{z}$ first and checking $\mathbf{r}_0$ first (Cycles). }
	\label{Table:checkref}
\begin{tabular}{c c c c}
%\specialrule{1pt}{0pt}{0pt}
%Scheme     & Round3 C REF(check $\mathbf{z}$ first) & Round3 C REF(check $\mathbf{r}_0$ first) & Imp.(\%) \\ \hline
%%%7 8 -> 10 11 12 13 14
\specialrule{1pt}{0pt}{0pt}

\multirow{2}{*}{\centering Scheme} & \multicolumn{2}{c}{Round3 C REF} & \multirow{2}{*}{Imp. (\%)} \\
\cmidrule(lr){2-3}
& \multicolumn{1}{c}{(check $\mathbf{z}$ first)} & \multicolumn{1}{c}{(check $\mathbf{r}_0$ first)} & \\
\hline

%Scheme & \multicolumn{2}{c}{Round3 C REF} & Imp. (\%) \\
%\cline{2-3}
%& \multicolumn{1}{c}{(check $\mathbf{z}$ first)} & %\multicolumn{1}{c}{(check $\mathbf{r}_0$ first)} & \\
%\hline

Dilithium2 & 992696                 & 972244                  & 2.06\%   \\ \hline
Dilithium3 & 1670374                & 1627560                 & 2.56\%   \\ \hline
Dilithium5 & 2088720                & 2026818                 & 2.96\%   \\ \specialrule{1pt}{0pt}{0pt}
\end{tabular}
	
\end{table}

%% file: PSPMref.tex
\captionsetup[table]{labelfont={bf}}
\begin{table}[htbp]
    \centering
	\renewcommand\arraystretch{1}
	\caption{Comparative performance of improved PSPM and original PSPM \cite{DBLP:conf/acsac/ZhengHSXZ22} (Cycles). }
	\label{Table:pspmwithcheck}
\begin{tabular}{c c c c}
\specialrule{1.0pt}{0pt}{0pt}
Scheme     & Sign (Original PSPM) & Sign (Improved PSPM) & Imp.(\%) \\ \hline
Dilithium2 &         670970            &     636326              & 5.16\%   \\ \hline
Dilithium3 &         1171086            &  1101330             & 6.00\%   \\ \hline
Dilithium5 &     1491124            &        1415452           & 5.07\%   \\ \specialrule{1.0pt}{0pt}{0pt}
\end{tabular}
	
\end{table}

%% file: checkr0z.tex
\begin{algorithm}[H]
  \caption{A parallel index-based polynomial multiplication algorithm with early evaluating $\mathbf{r}_0$ and $\mathbf{z}$ for Dilithium2}
  \label{algo-checkr0z}
  \small
  \begin{algorithmic}[1]
    \Statex \textbf{Input:}  $(c, \mathbf{s}, \mathbf{e}, \mathbf{y}, \mathbf{w})$, where $\mathbf{s} = [s^{(0)}, \cdots, s^{(l-1)}]^T \in \mathcal{R}_{q}^l, \mathbf{y} \in \mathcal{R}_{q}^l, \mathbf{e} \in \mathcal{R}_{q}^k, \mathbf{w} \in \mathcal{R}_{q}^k$ , every $s^{(j)}=\sum_{i=0}^{n-1} s^{(j)}_{i} \cdot x^{i} \in \mathcal{R}_{q}$, $y^{(j)}=\sum_{i=0}^{n-1} y^{(j)}_{i} \cdot y^{i} \in \mathcal{R}_{q}$, $e^{(j)}=\sum_{i=0}^{n-1} e^{(j)}_{i} \cdot e^{i} \in \mathcal{R}_{q}$, $w^{(j)}=\sum_{i=0}^{n-1} w^{(j)}_{i} \cdot w^{i} \in \mathcal{R}_{q}$, and $c=\sum_{i=0}^{n-1} c_{i} \cdot x^{i} \in B_{\tau}$
    \Statex \textbf{Output:}  $\mathbf{z} = c \cdot \mathbf{s} + \mathbf{y}= [z^{(0)}, \cdots, z^{(l-1)}]^T \in \mathcal{R}_{q}^l$, where $z^{(j)}=c \cdot s^{(j)} + y^{(j)}= \sum_{i=0}^{n-1} z^{(j)}_{i} \cdot x^{i} \in \mathcal{R}_{q}$, $\mathbf{r} = \mathbf{w} - c \cdot \mathbf{e}= [r^{(0)}, \cdots, r^{(k-1)}]^T \in \mathcal{R}_{q}^k$, where $r^{(j)}=w^{(j)} - c \cdot e^{(j)}= \sum_{i=0}^{n-1} r^{(j)}_{i} \cdot x^{i} \in \mathcal{R}_{q}$
    
    \For {$i \in \{0,1,\cdots, n-1\}$}
      \State $m_i := 0$
      \State $v_i := 0$
      \State $v_{i-n} := 0$
      \For {$j \in (0, 1, \cdots, l-1)$}
        \State $v_{i}:=v_{i} \cdot M+\left(U+s_{i}^{(j)}\right)$
        \State $v_{i-n}:=v_{i-n} \cdot M+\left(U-s_{i}^{(j)}\right)$
      \EndFor
      \For {$j \in (0, 1, \cdots, k-1)$}
        \State $v_{i}:=v_{i} \cdot M+\left(U+e_{i}^{(j)}\right)$
        \State $v_{i-n}:=v_{i-n} \cdot M+\left(U-e_{i}^{(j)}\right)$
      \EndFor
    \EndFor
    \State $\gamma:=2 U \cdot \frac{M^{l+k}-1}{M-1}$
    \For {$i \in \{0,1,\cdots, n-1\}$}
      \If {$c_i = 1$}
        \For {$j \in \{0,1,\cdots, n-1\}$}
          \State $m_{j} := m_{j} + v_{j-i}$
        \EndFor
      \EndIf
      \If {$c_i = -1$}
        \For {$j \in \{0,1,\cdots, n-1\}$}
          \State $m_{j} := m_{j} + (\gamma - v_{j-i})$
        \EndFor
      \EndIf
    \EndFor
    \For {$i \in \{0,1,\cdots, n-1\}$}
      \State $t := m_i$
      \For {$j \in (0, 1, \cdots, k-1)$}
        \State	$r_{i}^{(k-1-j)}:= (t \ \mathrm{mod} \ {M})-\tau U\pmod{q}$
                    \State  $r_{i}^{(k-1-j)}:= w_{i}^{(k-1-j)} - r_{i}^{(k-1-j)}$
                    \State  $r_{i}^{(k-1-j)}:=$\textbf{LowBits}$_q$$(r_{i}^{(k-1-j)},2\gamma_2)$
                    \If{$|r_{i}^{(k-1-j)}|>=\gamma_2-\beta$}
                        Restart signature process.
                    \EndIf
        \State $t:=\lfloor t / M\rfloor$
         \EndFor
      \For {$j \in (0, 1, \cdots, l-1)$}
        \State	$z_{i}^{(l-1-j)}:= (t \ \mathrm{mod} \ {M})-\tau U\pmod{q}$
                    \State  $z_{i}^{(l-1-j)}:= z_{i}^{(l-1-j)} + y_{i}^{(l-1-j)}$
                    \If{$|z_{i}^{(l-1-j)}|>=\gamma_1-\beta$}
                        Restart signature process.
                    \EndIf
        \State $t:=\lfloor t / M\rfloor$
      \EndFor
    \EndFor
    \State \textbf{return} $\mathbf{z} = [z^{(0)}, \cdots, z^{(l-1)}]^T$,$\mathbf{r} = [r^{(0)}, \cdots, r^{(k-1)}]^T$
  \end{algorithmic}
\end{algorithm}

%% file: tailored_reduction2.tex
\begin{algorithm}[H]
  \caption{Tailored reduction for the Dilithium prime $q = 2^{23} - 2^{13} + 1$}
  \label{algo-specialreduce2}
  \small
  \begin{algorithmic}[1]
    \Require $ -2^{40} < z \leq 2^{40}, q = 2^{23} - 2^{13} + 1$
    \Ensure $r=z (\bmod q), -2^{31} < r < 2^{31}$
    \State $p_{1}=\lfloor \frac{z}{2^{23}} \rfloor$
    \State $r= z-qp_{1}$
  \end{algorithmic}
\end{algorithm}

%% file: montreduction.tex
\begin{algorithm}[H]
  \caption{Signed Montgomery reduction for 32-bit $q$ \cite{DBLP:journals/iacr/Seiler18}}
  \label{algo-montreduction}
  \small
  \begin{algorithmic}[1]
    \Require $0<q<2^{31}$ odd, $-2^{31} q \leq z=z_1 2^{32}+z_0<2^{31} q$ where $0 \leq z_0<2^{32}$
    \Ensure $r^{\prime} \equiv \beta^{-1} z(\bmod q),-q<r^{\prime}<q$
    \State $m \leftarrow z_0 q^{-1} \bmod^{\pm} 2^{32}$ \Comment{signed low product, $q^{-1}$ precomputed}
    \State $t_1 \leftarrow\left\lfloor\frac{m q}{\beta}\right\rfloor$ \Comment signed high product
    \State $r^{\prime} \leftarrow z_1-t_1$
  \end{algorithmic}
\end{algorithm}

%% file: percentage_of_used_functions_in_keygen_sign_and_verify.tex
\captionsetup[table]{labelfont={bf}}
\begin{table}[htb]
    \caption{Percentages of used functions in Keygen, Signature and Verification. }
    \setlength{\tabcolsep}{1.2mm}
    \renewcommand\arraystretch{1.2}
    \begin{center}
    \begin{tabular}{ c c c c }
    \specialrule{1pt}{0pt}{0pt}
     Functions & Keygen & Sign& Verify\\\hline
     \textsf{montgomery\_reduce}            & 38.24\%                    & 23.02\%                  & 16.04\%                    \\\hline
     \textsf{KeccakF1600\_StatePermute}     & 17.68\%                    & 38.84\%                  & 42.99\%                    \\\hline
     \textsf{invntt\_tomont}                & 17.48\%                    & 6.28\%                   & 5.22\%                     \\\hline
     \textsf{ntt}                           & 6.94\%                     & 9.30\%                   & 3.35\%                     \\\hline
     \textsf{poly\_pointwise\_montgomery}   & 4.96\%                     & 1.86\%                   & 2.22\%                   \\ \specialrule{1pt}{0pt}{0pt}
\end{tabular}
\label{Table:functionsused}
\end{center}
\end{table}

%% file: tailored_reduction3instr.tex
\begin{algorithm}[H]
  \caption{2-instruction Tailored reduction using AVX512IFMA}
  \label{algo-tailord}
  \small
  \begin{algorithmic}[1]
    \Statex \textbf{Input:}  A 40-bit signed integer $-2^{40} < z \leq 2^{40}$
    \Statex \textbf{Output:} $r=z (\bmod q), -2^{31} < r < 2^{31}$ 
        \State  \textbf{vpsrlq}  $23,z,r$ \Comment{$\frac{z}{2^{23}}$}
        \State \textbf{vpmadd52luq}  $-q,z,r$ \Comment{$z-\frac{z}{2^{23}} \cdot q$}
    \State \textbf{return} $r$
  \end{algorithmic}
\end{algorithm}

%% file: tailoredreductionAVX-512.tex
\begin{algorithm}[H]
  \caption{3-instruction Tailored reduction using AVX512}
  \label{algo-tailord3}
  \small
  \begin{algorithmic}[1]
    \Statex \textbf{Input:}  A 40-bit signed integer $-2^{40} < z \leq 2^{40}$
    \Statex \textbf{Output:} $r=z (\bmod q), -2^{31} < r < 2^{31}$ 
        \State  \textbf{vpsrlq}  $23,z,r$ \Comment{$\frac{z}{2^{23}}$}
        \State \textbf{vpmuldq}  $q,z,t$ \Comment{t = $\frac{z}{2^{23}} \cdot q$}
        \State \textbf{vpsubq}  $t,z,r$ \Comment{$r = z - t$}
    \State \textbf{return} $r$
  \end{algorithmic}
\end{algorithm}

%% file: MontgomeryreductionAVX-512.tex
\begin{algorithm}[H]
  \caption{4-instruction Montgomery reduction using AVX512 \cite{avanzi2022dilithium}}
  \label{algo-mont512}
  \small
  \begin{algorithmic}[1]
    \Statex \textbf{Input:}  A signed integer $-2^{31}q < z \leq 2^{31}q$
    \Statex \textbf{Output:} $r'=2^{-32}z (\bmod q), -q < r' < q$ 
        \State  \textbf{vpmuldq}  $q^{-1},z,m$ \Comment{$m = z \bmod 2^{32} \cdot q^{-1}$}
        \State \textbf{vpmuldq}  $q,m,t$ \Comment{$t = m \bmod 2^{32} \cdot q$}
        \State \textbf{vpsubq}  $t,z,r'$ \Comment{$r' = z - t $}
        \State \textbf{vpsrlq}  $32,r',r'$ \Comment{$r' = \frac{r'}{2^{32}} $}
    \State \textbf{return} $r'$
  \end{algorithmic}
\end{algorithm}

%% file: preandevaluate.tex
\captionsetup[table]{labelfont={bf}}
\begin{table}[!ht]
   \caption{Performance comparison of 32-bit version and 64-bit version PSPM.}
	\label{Tab:preandeva}
\begin{tabularx}{\linewidth}{@{}l*{3}{>{\centering\arraybackslash}X}@{}}

\specialrule{1.0pt}{0pt}{0pt}
   Operation  &  Scheme & 32-bit version (Cycles) & 64-bit version (Cycles) \\ \hline
Preparing $ \mathbf{s}$ $\mathbf{e}$                & Dilithium2 & 264            & 536             \\ \hline
\multirow{2}{*}{Preparing $\mathbf{s}$}  & Dilithium3 & 320             & 378            \\ \cline{2-4} 
                               & Dilithium5 & 440            & 918           \\ \hline
\multirow{2}{*}{Preparing $\mathbf{e}$}  & Dilithium3 &  338            &  448            \\ \cline{2-4} 
                               & Dilithium5 & 486            & 1052           \\ \hline
Evaluating $c\mathbf{s}$ $c\mathbf{e}$              & Dilithium2 & 5358           & 5800           \\ \hline
\multirow{2}{*}{Evaluating $c\mathbf{s}$} & Dilithium3 & 3760           & 5160           \\ \cline{2-4} 
                               & Dilithium5 & 4794          & 6502          \\ \hline
\multirow{2}{*}{Evaluating $c\mathbf{e}$} & Dilithium3 & 6280          & 7846          \\ \cline{2-4} 
                               & Dilithium5 & 3156           & 4892          \\ 
\specialrule{1.0pt}{0pt}{0pt}

\end{tabularx}
 
\end{table}

%% file: performance_of_PSPM.tex
\captionsetup[table]{labelfont={bf}}
\begin{table}[htbp]
  \centering
 \caption{Performance of $c\cdot\mathbf{s}$ and
$c\cdot\mathbf{e}$ in Dilithium3 (Cycles). }
	\label{Table:pmperf}
  \begin{tabularx}{\linewidth}{@{}>{\centering\arraybackslash}X *{3}{>{\centering\arraybackslash}X}@{}}
   \specialrule{1.0pt}{0pt}{0pt}

                         & AVX2  \cite{avanzi2022dilithium}  & AVX-512 \\ \hline
$c\cdot\mathbf{s}$ (PSPM) & 6748  & 2556   \\\hline
$c\cdot\mathbf{s}$ (NTT)  & 14636 & 6740   \\\hline
$c\cdot\mathbf{e}$ (PSPM) & 8358  & 2560   \\\hline
$c\cdot\mathbf{e}$ (NTT)  & 16010 & 7480   
	\\ \specialrule{1.0pt}{0pt}{0pt}

    \end{tabularx}

\end{table}

%% file: PSPMperformance.tex
\captionsetup[table]{labelfont={bf}}
\begin{table}[htbp]
  \centering
  	\caption{Performance of Signing Procedure with Improved PSPM (Cycles). }
	\label{Table:pspmperf}
  \begin{tabularx}{\linewidth}{@{}>{\centering\arraybackslash}X *{3}{>{\centering\arraybackslash}X}@{}}

%	\renewcommand\arraystretch{1}
%	\begin{tabular}{*{4}{c}}
		\specialrule{1.0pt}{0pt}{0pt}
%		{}
%			& {$\mathrm{AVX2}$}
%			& {$\mathrm{AVX2(improved \; PSPM)}$}
%            & {$\mathrm{Speedup}$}
%			\\
%			\hline

\multirow{2}{*}{}
	& \multirow{2}{*}{$\mathrm{AVX2}$} & $\mathrm{AVX2}$ & \multirow{2}{*}{$\mathrm{Speedup}$} \\
	& & $\mathrm{(PSPM-TEE)}$& \\
	\hline

		{Dilithium2}
			& { 251050}
			& {231766}
            & {7.7\%}
			\\ \hline
		{Dilithium3}
			& {406248}
			& {393454}
            & {3.1\%}
			\\ \hline
		{Dilithium5}
			& {516200}
			& {501304}
            & {2.9\%}
			\\ \specialrule{1.0pt}{0pt}{0pt}
    \end{tabularx}

\end{table}

%% file: vecfunction.tex
\captionsetup[table]{labelfont={bf}}
\begin{table}[htbp]
    \centering
    \caption{Experimental results of vectorization functions for Dilithium (Cycles). }
	\label{Table:vecfunc}  
%    \begin{tabularx}{\linewidth}{>{\centering\arraybackslash}p{4cm}*{3}{>{\centering\arraybackslash}X}}

\begin{tabularx}{1\linewidth}{>{\centering\arraybackslash}p{3cm}*{3}{>{\centering\arraybackslash}X}}

\specialrule{1pt}{0pt}{0pt}
Function                                       & Vectorization & Cycles         & Speedup       \\ \hline
\multirow{3}{*}{\textsf{Poly\_uniform}}                  & 1-way         & 5784 & 1.00× \\ \cline{2-4} 
                                                & 4-way         & 19488 &  2.97×  \\ \cline{2-4} 
                                                & 8-way         &  13450  &  4.30×  \\ \hline
\multirow{3}{*}{\textsf{Poly\_uniform\_eta}}    & 1-way         &  30158 &  1.00× \\ \cline{2-4} 
                                                & 4-way         &   17858  & 1.69×          \\ \cline{2-4} 
                                                & 8-way         &  9054   & 3.33×          \\ \hline
\multirow{3}{*}{\textsf{Poly\_uniform\_gamma1}} & 1-way         &  48148 &  1.00×  \\ \cline{2-4} 
                                                & 4-way         &  24594  & 1.95×          \\ \cline{2-4} 
                                                & 8-way         & 12094 & 3.98×          \\ \hline
\multirow{2}{*}{\texttt{SHAKE-256} (sequential)}                   & 1-way         &  1300             & 1.00×\\ \cline{2-4} 
                                                & 5-way         &  918              & 1.41× \\ \hline        
\multirow{3}{*}{\texttt{SHAKE-256} (parallel)}                   & 1-way         &  5934             & 1.00×\\ \cline{2-4} 
                                                & 4-way         & 2896             & 2.05× \\ \cline{2-4} 
                                                & 8-way         &  1014              & 5.85× \\ \hline                                          

\multirow{3}{*}{\texttt{SHAKE-128} (parallel)}                   & 1-way         &  6126             &  1.00×  \\ \cline{2-4} 
                                                & 4-way         & 3006             & 2.04× \\ \cline{2-4} 
                                                & 8-way         &  1114              & 5.50× \\ \hline                                                    
\multirow{3}{*}{\textsf{rej\_uniform}}                   & 1-way         & 450            &  1.00×  \\ \cline{2-4} 
                                                & 8-way         & 230   & 1.96×          \\ \cline{2-4} 
                                                & 16-way         & 80             & 5.63×          \\ \hline
\multirow{3}{*}{\textsf{rej\_eta}}              & 1-way         &      1122           &     1.00×            \\ \cline{2-4} 
                                                & 8-way         &      322           & 3.48×             \\ \cline{2-4} 
                                                & 16-way         &    230               &   4.88× \\ \hline                                                
\multirow{3}{*}{\texttt{NTT}}                            & 1-way         & 6896           & \textsf{1.00×} \\ \cline{2-4} 
                                                & 4-way         & 1326           & 5.00×          \\ \cline{2-4} 
                                                & 16-way         & 494            & 13.95×         \\ \hline
\multirow{3}{*}{$\texttt{NTT}^{-1}$}                 & 1-way         & 9438           &  1.00× \\ \cline{2-4} 
                                                & 4-way         & 1090           & 8.66×          \\ \cline{2-4} 
                                                & 16-way         & 526            & 17.94×         \\ \hline
\multirow{3}{*}{\textsf{poly\_pointwise}}                & 1-way         & 1374           &  1.00× \\ \cline{2-4} 
                                                & 4-way         & 146            & 9.41×          \\ \cline{2-4} 
                                                & 16-way         & 124            & 11.08×         \\ \hline
\multirow{3}{*}{\textsf{polyz\_unpack}}                  & 1-way         & 962            & 1.00× \\ \cline{2-4} 
                                                & 32-way         &      -        &       -      \\ \cline{2-4} 
                                                & 64-way         & 32             & 30×            \\ \hline
\multirow{3}{*}{\textsf{polyw1\_pack}}                   & 1-way         & 32             &  1.00×  \\ \cline{2-4} 
                                                & 8-way         & 32             & 1.00× \\ \cline{2-4} 
                                                & 16-way         & 16             & 2.00× \\ 
             \specialrule{1pt}{0pt}{0pt}
\end{tabularx}

\end{table}

%% file: schemeperformance.tex
\captionsetup[table]{labelfont={bf}}
\begin{table}[htbp]
\centering
\caption{Execution times (in Cycles) of implementation of Round3 Dilithium2, Dilithium3 and Dilithium5 on an Intel Core i7-11700F processor.}
\label{tab:my-table}
\resizebox{\columnwidth}{!}{%
\begin{tabular}{c c c c c c c}
\specialrule{1.0pt}{0pt}{0pt}
\multirow{2}{*}{Scheme}     & \multirow{2}{*}{Operation} & C\cite{avanzi2022dilithium}       & AVX2\cite{avanzi2022dilithium}   & \multicolumn{3}{c}{AVX-512}                                                  \\
                            &                            & Cycles  & Cycles & Cycles                      & Speedup vs C               & Speedup vs AVX2 \\ \hline
\multirow{3}{*}{Dilithium2} & KeyGen                     & 266772  & 106000  & \multicolumn{1}{c }{47168}  & \multicolumn{1}{c }{82.3\%} & 55.5\%           \\
                            & Sign                       & 1033894
  & 251050
 & \multicolumn{1}{c }{125554
}  & \multicolumn{1}{c }{87.9\%} & 50.0\%           \\
                            & Verify                     & 298384
  & 107338  & \multicolumn{1}{c }{48320}  & \multicolumn{1}{c }{83.8\%} & 55.0\%             \\ \hline
\multirow{3}{*}{Dilithium3} & KeyGen                     & 503306
  & 246988
 & \multicolumn{1}{c }{86114
}  & \multicolumn{1}{c }{82.9\%} & 65.1\%           \\
                            & Sign                       & 1699294
  & 406248
 & \multicolumn{1}{c }{191946}  & \multicolumn{1}{c }{88.7\%} & 52.8\%           \\
                            & Verify                     & 478660
  & 174218
  & \multicolumn{1}{c }{76256
}  & \multicolumn{1}{c }{84.1\%} & 56.2\%           \\ \hline
\multirow{3}{*}{Dilithium5} & KeyGen                     & 725802
  & 286534
 & \multicolumn{1}{c }{118568
}  & \multicolumn{1}{c }{83.7\%} & 58.6\%           \\
                            & Sign                       & 2111234
 & 516200
 & \multicolumn{1}{c }{223776
} & \multicolumn{1}{c }{89.4\%} & 56.6\%           \\
                            & Verify                     & 770794
  & 275894
 & \multicolumn{1}{c }{114412
}  & \multicolumn{1}{c }{85.2\%} & 58.5\%             \\ \specialrule{1.0pt}{0pt}{0pt}
\end{tabular}%
}
\label{tab:scheme}
\end{table}

%% file: AVX2optimizedperformance.tex
\captionsetup[table]{labelfont={bf}}
\begin{table}[htb]
    \caption{Performance comparison in Signing procedure (Cycles).}
    \setlength{\tabcolsep}{1.2mm}
    \renewcommand\arraystretch{1.2}
    \begin{center}
    \begin{tabular}{ c c c c }
    \specialrule{1.0pt}{0pt}{0pt}
     Scheme & AVX2 \cite{avanzi2022dilithium} & AVX2 (Our work)& Speedup\\\hline
     Dilithium2            & 251050                    &231410                  & 7.8\%                  \\\hline
     Dilithium3     & 406248                   & 392436                  & 3.4\%                    \\\hline
    Dilithium5                & 516200                   & 500882                  &3.0\%                     \\ \specialrule{1.0pt}{0pt}{0pt}
\end{tabular}
\label{Table:AVX2optimized}
\end{center}
\end{table}

%% file: standardperformance.tex
\captionsetup[table]{labelfont={bf}}
\begin{table}[htbp]
\centering
\caption{Execution times (in Cycles) of implementation of ML-DSA \cite{dilidraft2023} on an Intel Core i7-11700F processor.}
\label{tab:my-table}
\resizebox{\columnwidth}{!}{%
\begin{tabular}{c c c c c c c}
\specialrule{1.0pt}{0pt}{0pt}
\multirow{2}{*}{Scheme}     & \multirow{2}{*}{Operation} & C\cite{dilidraft2023}       & AVX2\cite{dilidraft2023}   & \multicolumn{3}{c}{AVX-512}                                                  \\
                            &                            & Cycles  & Cycles & Cycles                      & Speedup vs C               & Speedup vs AVX2 \\ \hline
\multirow{3}{*}{ML-DSA-44} & KeyGen                     & 299120  & 84270  & \multicolumn{1}{c }{47760}  & \multicolumn{1}{c }{84.0\%} & 43.2\%           \\
                            & Sign                       & 1068726
  & 194320
 & \multicolumn{1}{c }{123160
}  & \multicolumn{1}{c }{88.4\%} & 36.6\%           \\
                            & Verify                     & 328798
  & 90314  & \multicolumn{1}{c }{49396}  & \multicolumn{1}{c }{84.9\%} & 45.3\%             \\ \hline
\multirow{3}{*}{ML-DSA-65} & KeyGen                     & 558152
  & 144114
 & \multicolumn{1}{c }{87378
}  & \multicolumn{1}{c }{84.3\%} & 39.3\%           \\
                            & Sign                       & 1804702
  & 327856
 & \multicolumn{1}{c }{191152}  & \multicolumn{1}{c }{89.4\%} & 41.6\%           \\
                            & Verify                     & 533314
  & 145990
  & \multicolumn{1}{c }{78008
}  & \multicolumn{1}{c }{85.3\%} & 46.5\%           \\ \hline
\multirow{3}{*}{ML-DSA-87} & KeyGen                     & 818616
  & 224466
 & \multicolumn{1}{c }{122032
}  & \multicolumn{1}{c }{85.1\%} & 45.6\%           \\
                            & Sign                       & 2208464
 & 402276
 & \multicolumn{1}{c }{226272
} & \multicolumn{1}{c }{89.7\%} & 43.7\%           \\
                            & Verify                     & 856658
  & 227138
 & \multicolumn{1}{c }{119428
}  & \multicolumn{1}{c }{86.1\%} & 47.4\%             \\ \specialrule{1.0pt}{0pt}{0pt}
\end{tabular}%
}
\label{tab:standscheme}
\end{table}

%% file: checkz.tex
\begin{algorithm}[htpb]
  \caption{A parallel index-based polynomial multiplication algorithm with early evaluating $\mathbf{z}$ for Dilithium3/5}
  \label{algo-checkz}
  \small
  \begin{algorithmic}[1]
    \Statex \textbf{Input:}  $(c, \mathbf{s}, \mathbf{y})$, where $\mathbf{s} = [s^{(0)}, \cdots, s^{(l-1)}]^T \in \mathcal{R}_{q}^l, \mathbf{y} \in \mathcal{R}_{q}^l$ , every $s^{(j)}=\sum_{i=0}^{n-1} s^{(j)}_{i} \cdot x^{i} \in \mathcal{R}_{q}$, $y^{(j)}=\sum_{i=0}^{n-1} y^{(j)}_{i} \cdot y^{i} \in \mathcal{R}_{q}$, and $c=\sum_{i=0}^{n-1} c_{i} \cdot x^{i} \in B_{\tau}$
    \Statex \textbf{Output:}  $\mathbf{z} = c \cdot \mathbf{s} + \mathbf{y}= [z^{(0)}, \cdots, z^{(l-1)}]^T \in \mathcal{R}_{q}^l$, where $z^{(j)}=c \cdot s^{(j)} + y^{(j)}= \sum_{i=0}^{n-1} z^{(j)}_{i} \cdot x^{i} \in \mathcal{R}_{q}$
    
    \For {$i \in \{0,1,\cdots, n-1\}$}
      \State $m_i := 0$
      \State $v_i := 0$
      \State $v_{i-n} := 0$
      \For {$j \in (0, 1, \cdots, l-1)$}
        \State $v_{i}:=v_{i} \cdot M+\left(U+s_{i}^{(j)}\right)$
        \State $v_{i-n}:=v_{i-n} \cdot M+\left(U-s_{i}^{(j)}\right)$
      \EndFor
    \EndFor
    \State $\gamma:=2 U \cdot \frac{M^{l}-1}{M-1}$
    \For {$i \in \{0,1,\cdots, n-1\}$}
      \If {$c_i = 1$}
        \For {$j \in \{0,1,\cdots, n-1\}$}
          \State $m_{j} := m_{j} + v_{j-i}$
        \EndFor
      \EndIf
      \If {$c_i = -1$}
        \For {$j \in \{0,1,\cdots, n-1\}$}
          \State $m_{j} := m_{j} + (\gamma - v_{j-i})$
        \EndFor
      \EndIf
    \EndFor
    \For {$i \in \{0,1,\cdots, n-1\}$}
      \State $t := m_i$
      \For {$j \in (0, 1, \cdots, l-1)$}
        \State	$z_{i}^{(l-1-j)}:= (t \ \mathrm{mod} \ {M})-\tau U\pmod{q}$
                    \State  $z_{i}^{(l-1-j)}:= z_{i}^{(l-1-j)} + y_{i}^{(l-1-j)}$
                    \If{$|z_{i}^{(l-1-j)}|>=\gamma_1-\beta$}
                        Restart signature process.
                    \EndIf
        \State $t:=\lfloor t / M\rfloor$
      \EndFor
    \EndFor
    \State \textbf{return} $\mathbf{z} = [z^{(0)}, \cdots, z^{(l-1)}]^T$
  \end{algorithmic}
\end{algorithm}

%% file: checkr0.tex
\begin{algorithm}[htp]
  \caption{A parallel index-based polynomial multiplication algorithm with early evaluating $\mathbf{r}_0$ for Dilithium3/5}
  \label{algo-checkr0}
  \small
  \begin{algorithmic}[1]
    \Statex \textbf{Input:}  $(c, \mathbf{e}, \mathbf{w})$, where $\mathbf{e} = [e^{(0)}, \cdots, e^{(k-1)}]^T \in \mathcal{R}_{q}^k, \mathbf{w} \in \mathcal{R}_{q}^k$, every $e^{(j)}=\sum_{i=0}^{n-1} e^{(j)}_{i} \cdot x^{i} \in \mathcal{R}_{q}$, $w^{(j)}=\sum_{i=0}^{n-1} w^{(j)}_{i} \cdot w^{i} \in \mathcal{R}_{q}$, and $c=\sum_{i=0}^{n-1} c_{i} \cdot x^{i} \in B_{\tau}$
    \Statex \textbf{Output:}  $\mathbf{r} = c \cdot \mathbf{e} - \mathbf{w}= [r^{(0)}, \cdots, r^{(k-1)}]^T \in \mathcal{R}_{q}^k$, where $r^{(j)}=c \cdot e^{(j)} - w^{(j)}= \sum_{i=0}^{n-1} r^{(j)}_{i} \cdot x^{i} \in \mathcal{R}_{q}$
    
    \For {$i \in \{0,1,\cdots, n-1\}$}
      \State $m_i := 0$
      \State $v_i := 0$
      \State $v_{i-n} := 0$
      \For {$j \in (0, 1, \cdots, k-1)$}
        \State $v_{i}:=v_{i} \cdot M+\left(U+e_{i}^{(j)}\right)$
        \State $v_{i-n}:=v_{i-n} \cdot M+\left(U-e_{i}^{(j)}\right)$
      \EndFor
    \EndFor
    \State $\gamma:=2 U \cdot \frac{M^{k}-1}{M-1}$
    \For {$i \in \{0,1,\cdots, n-1\}$}
      \If {$c_i = 1$}
        \For {$j \in \{0,1,\cdots, n-1\}$}
          \State $m_{j} := m_{j} + v_{j-i}$
        \EndFor
      \EndIf
      \If {$c_i = -1$}
        \For {$j \in \{0,1,\cdots, n-1\}$}
          \State $m_{j} := m_{j} + (\gamma - v_{j-i})$
        \EndFor
      \EndIf
    \EndFor
    \For {$i \in \{0,1,\cdots, n-1\}$}
      \State $t := m_i$
      \For {$j \in (0, 1, \cdots, k-1)$}
        \State	$r_{i}^{(k-1-j)}:= (t \ \mathrm{mod} \ {M})-\tau U\pmod{q}$
                    \State  $r_{i}^{(k-1-j)}:= m_{i}^{(k-1-j)} - r_{i}^{(k-1-j)}$
                    \State  $r_{i}^{(k-1-j)}:=$\textbf{LowBits}$_q$$(r_{i}^{(k-1-j)},2\gamma_2)$
                    \If{$|r_{i}^{(k-1-j)}|>=\gamma_2-\beta$}
                        Restart signature process.
                    \EndIf
        \State $t:=\lfloor t / M\rfloor$
      \EndFor
    \EndFor
    \State \textbf{return} $\mathbf{r} = [r^{(0)}, \cdots, r^{(k-1)}]^T$
  \end{algorithmic}
\end{algorithm}

%% file: parallel_parameter.tex
\captionsetup[table]{labelfont={bf}}
\begin{table}[htbp]
    \caption{Parallel Parameters of Dilithium.}
    \setlength{\tabcolsep}{2mm}
    \renewcommand\arraystretch{1.2}
    \begin{center}
        \begin{tabular}{ c|c|c|c|c|c|c }
        \specialrule{1pt}{0pt}{0pt}
        Scheme & Operation & $\tau$ & $U$ & $2 \tau U$ & $\mathrm{M}$ & $r$ \\ \hline\hline
        
        \multirow{4}{*}{Dilithium2} & $c\mathbf{s}_{1}$ & 39 & 2 & 156 & $2^{8}$ & 4 \\\cline{2-7} 
                                    & $c\mathbf{s}_{2}$ & 39 & 2 & 156 & $2^{8}$ & 4 \\\cline{2-7}
                                    & $c\mathbf{t}_{0}$ & 39 & $2^{12}$ & 319488 & $2^{19}$ & 4 \\ \cline{2-7}
                                    & $c\mathbf{t}_{1}$ & 39 & $2^{10}$ & 79872 & $2^{17}$ & 4 \\ \hline
        
        \multirow{4}{*}{Dilithium3} & $c\mathbf{s}_{1}$ & 49 & 4 & 392 & $2^{9}$ & 5 \\\cline{2-7}
                                    & $c\mathbf{s}_{2}$ & 49 & 4 & 392 & $2^{9}$ & 6 \\\cline{2-7}
                                    & $c\mathbf{t}_{0}$ & 49 & $2^{12}$ & 401408 & $2^{19}$ & 6 \\\cline{2-7}
                                    & $c\mathbf{t}_{1}$ & 49 & $2^{10}$ & 100352 & $2^{17}$ & 6 \\\hline
                                    
        \multirow{4}{*}{Dilithium5} & $c\mathbf{s}_{1}$ & 60 & 2 & 240 & $2^{8}$ & 7 \\\cline{2-7}
                                    & $c\mathbf{s}_{2}$ & 60 & 2 & 240 & $2^{8}$ & 8 \\\cline{2-7}
                                    & $c\mathbf{t}_{0}$ & 60 & $2^{12}$ & 491520 & $2^{19}$ & 8 \\\cline{2-7}
                                    & $c\mathbf{t}_{1}$ & 60 & $2^{10}$ & 122880 & $2^{17}$ & 8 \\\specialrule{1pt}{0pt}{0pt}
        \end{tabular}
    \label{tab-parapram}
    \end{center}
\end{table}